# Achieving the Secrecy Capacity of Wiretap Channels Using Polar Codes

Hessam Mahdavifar, *Student Member, IEEE*, and Alexander Vardy, *Fellow, IEEE*

*Abstract*— Suppose that Alice wishes to send messages to Bob through a communication channel $C_1$, but her transmissions also reach an eavesdropper Eve through another channel $C_2$. This is the wiretap channel model introduced by Wyner in 1975. The goal is to design a coding scheme that makes it possible for Alice to communicate both *reliably* and *securely*. Reliability is measured in terms of Bob's probability of error in recovering the message, while security is measured in terms of the mutual information between the message and Eve's observations. Wyner showed that the situation is characterized by a single constant $C_s$, called the *secrecy capacity*, which has the following meaning: for all $\varepsilon > 0$, there exist coding schemes of rate $R \geqslant C_s - \varepsilon$ that asymptotically achieve both the reliability and the security objectives. However, his proof of this result is based upon a nonconstructive random-coding argument. To date, despite a considerable research effort, the only case where we know how to *construct* coding schemes that achieve secrecy capacity is when Eve's channel $C_2$ is an erasure channel, or a combinatorial variation thereof.

Polar codes were recently invented by Arıkan; they approach the capacity of symmetric binary-input discrete memoryless channels with low encoding and decoding complexity. In this paper, we use polar codes to construct a coding scheme that achieves the secrecy capacity for a wide range of wiretap channels. Our construction works for any instantiation of the wiretap channel model, as long as both $C_1$ and $C_2$ are symmetric and binary-input, and $C_2$ is degraded with respect to $C_1$. Moreover, we show how to modify our construction in order to provide *strong security*, in the sense defined by Maurer, while still operating at a rate that approaches the secrecy capacity. In this case, we cannot guarantee that the reliability condition will also be satisfied unless the main channel $C_1$ is noiseless, although we believe it can be always satisfied in practice.

## I. INTRODUCTION

THE notion of wiretap channels was introduced by Aaron Wyner [41] in 1975. In this setting, Alice wishes to send messages to Bob through a communication channel $C_1$, called the **main channel**, but her transmissions also reach an adversary Eve through another channel $C_2$, called the **wiretap channel**. This is illustrated in Figure 1, wherein $U$ denotes a $k$-bit message that Alice wishes to communicate to Bob. We think of $U$ as a random variable that takes values in $\{0,1\}^k$; unlike most papers on wiretap channels, we *do not assume anything* regarding the a priori distribution of $U$. While making use of auxiliary random bits, the encoder maps $U$ into a sequence $X$ of $n$ chan-

The material in this paper was presented in part at the IEEE International Symposium on Information Theory, Austin, Texas, June 2010.

Hessam Mahdavifar is with the Department of Electrical and Computer Engineering, University of California San Diego, La Jolla, CA 92093–0407, U.S.A. (e-mail: hessam@ucsd.edu).

Alexander Vardy is with the Department of Electrical and Computer Engineering, the Department of Computer Science and Engineering, and the Department of Mathematics, University of California San Diego, La Jolla, CA 92093–0407, U.S.A. (e-mail: avardy@ucsd.edu).

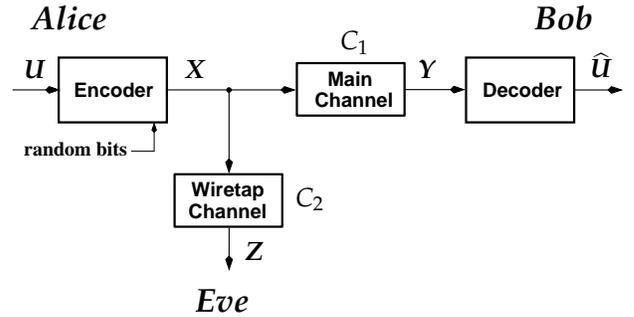

**Figure 1.** *Block diagram of a generic wiretap-channel system*

nel symbols. This sequence is transmitted across the main channel and the wiretap channel resulting in the corresponding channel outputs $Y$ and $Z$. Finally, the decoder maps $Y$ (deterministically) into an estimate $\widehat{U}$ of the original message.

The goal is to design a coding scheme — namely, an encoding algorithm and a decoding algorithm — that makes it possible to communicate both *reliably* and *securely*, as the message length $k$ tends to infinity. Reliability is measured in terms of the **probability of error** in recovering the message. Specifically, the objective is to satisfy the following

**Reliability Condition:** $\lim_{k \to \infty} \Pr\{\widehat{U} \neq U\} = 0$ (1)

where the probability is over all the relevant coin tosses in the system: in the generation of $U$, in the encoder, and in the main channel. Security is usually measured in terms of the **normalized mutual information** between the message $U$ and Eve's observations $Z$. Specifically, one is interested in encoding algorithms that satisfy the following

**Security Condition:** $\lim_{k \to \infty} \dfrac{I(U;Z)}{k} = 0$ (2)

Note that $I(U;Z)$ is equal to the difference between the a priori entropy $H(U)$ and the conditional entropy $H(U|Z)$. Thus, intuitively, (2) means that observing $Z$ does not provide much information about $U$ beyond what is available a priori, as compared to the message length $k$. Maurer argued in [27, 28] that the conventional notion of security (2) is much too weak. Indeed, it is easy to construct examples where $k^{1-\varepsilon}$ out of the $k$ message bits are disclosed to Eve, while still satisfying (2). This is clearly unacceptable. Thus Maurer introduced in [27] an alternative

**Strong Security Condition:** $\lim_{k \to \infty} I(U;Z) = 0$ (3)

Notice that both security conditions (2) and (3) are information-theoretic rather than computational: the adversary is assumed to be computationally unbounded, and security does not depend on computational hardness assumptions of any kind.

*A. Prior Work*

In 1975, Wyner [41] considered a special case of the system in Figure 1 where both $C_1$ and $C_2$ are discrete memoryless channels (DMCs) and, moreover, $C_2$ is degraded with respect to $C_1$. He proved that such a system is characterized by a single constant $C_s$, called the *secrecy capacity*, which has the following meaning. For all $\varepsilon > 0$, there exist coding schemes of information rate $R \geqslant C_s - \varepsilon$ that satisfy (1) and (2); conversely, it is not possible to satisfy both (1) and (2) at rates greater than $C_s$. Since 1975, Wyner's results have been extended to a variety of contexts, most notably Gaussian channels [22], general broadcast channels with confidential messages [11], and channels that impose a combinatorial (rather than probabilistic) constraint on the adversary [7, 32]. In fact, the literature on wiretap channels encompasses, by now, hundreds of papers.

However, the vast majority of this work relies on nonconstructive random-coding arguments to establish the main results. Such results show that *there exist* codes that achieve secrecy capacity, but are of little use if one's goal is to design specific polynomial-time encoding/decoding algorithms. To the best of our knowledge, *constructive solutions* to the wiretap-channel problem are available only in two special cases. The first special case is when the main channel is noiseless and the wiretap channel is the binary erasure channel (BEC). A coding scheme for this case, using LDPC codes for the BEC, was presented in [36, 38] and proved to achieve secrecy capacity. The other special case is when the adversary is constrained combinatorially: Eve can select to observe some $t$ out of the $n$ transmitted symbols, while the remaining $n - t$ symbols are erased. This situation, studied by Ozarow and Wyner in [32], may be regarded as a combinatorial variation of an erasure channel. Provably optimal coding schemes for this case can be constructed from MDS codes [40], or using extractors [7]. We observe, however, that even for the simple situation where $C_1$ is noiseless and $C_2$ is a binary symmetric channel, it is not known how to explicitly construct codes that achieve secrecy capacity.

We point out that a general method of coding for the wiretap channel, often referred to as *coset-coding* or *syndrome-coding*, is well known. This method goes back to the work of Wyner [32, 41], although it was significantly extended and generalized in [8, 9] and other papers. Assume, for simplicity, that the input alphabet of both $C_1$ and $C_2$ is binary. In this case, the coset-coding method utilizes two binary linear codes: an "outer" code $\mathbb{C}^*$ and an "inner" code $\mathbb{C}$, such that $\mathbb{C} \subset \mathbb{C}^*$ and the difference $\dim(\mathbb{C}^*) - \dim(\mathbb{C})$ between their dimensions is $k$. This condition implies that $\mathbb{C}^*$ can be partitioned into $2^k$ cosets of $\mathbb{C}$. A message $\boldsymbol{u} \in \{0,1\}^k$ is conveyed by Alice via the choice of one of these $2^k$ cosets, say $\boldsymbol{a} + \mathbb{C}$. What is transmitted by Alice is a vector $\boldsymbol{X}$ that is selected uniformly at random from $\boldsymbol{a} + \mathbb{C}$. Loosely speaking, the outer code $\mathbb{C}^*$ serves to correct the errors on the main channel, and thus ensures reliability, while the inner code $\mathbb{C}$, over which $\boldsymbol{X}$ is randomized, ensures security. The trouble is that it is not known how to *explicitly construct* a sequence of outer codes $\mathbb{C}^*$ and inner codes $\mathbb{C}$ that satisfy conditions (1) and (2) at a rate that approaches the secrecy capacity as $n \to \infty$. The remarkable work of Cohen and Zémor [8, 9] shows that a *random choice* of the inner code $\mathbb{C}$ suffices to achieve strong security, in a very general setting. Notably, the proof of this result in [8, 9] does *not* assume that the messages are uniformly random a priori. Still, to the best of our knowledge, the only cases where explicit constructions of $\mathbb{C}$ and $\mathbb{C}^*$ are known are those described in the foregoing paragraph (cf. [36, 38]).

*B. Our Contributions*

In this paper, we present a coding scheme that achieves the secrecy capacity of wiretap channels whenever $C_1$ and $C_2$ are binary-input symmetric DMCs and $C_2$ is degraded with respect to $C_1$. This is the situation originally studied by Wyner; it includes the important special case where $C_1$ and $C_2$ are arbitrary binary symmetric channels. We are able to satisfy the reliability and security conditions (1) and (2) with explicit polynomial-time encoding/decoding algorithms. In fact, the number of operations required for encoding and decoding is only $O(n \log n)$. Our construction is based upon key results in the literature on *polar codes*, recently invented by Arıkan [3].

It is proved in [3] that polar codes achieve the capacity of arbitrary binary-input symmetric DMCs, with low encoding and decoding complexity. The proof of this result is based on a phenomenon called *channel polarization*. Let

$$G = \begin{bmatrix} 1 & 0 \\ 1 & 1 \end{bmatrix} \quad (4)$$

and let $G^{\otimes m}$ denote the $m$-th Kronecker power of $G$. Let $W$ be a symmetric binary-input DMC, and let $\boldsymbol{V} = (V_1, V_2, \ldots, V_n)$ be a block of $n = 2^m$ bits chosen uniformly at random from $\{0,1\}^n$. Suppose $\boldsymbol{V}$ is encoded as $\boldsymbol{X} = \boldsymbol{V} P_n G^{\otimes m}$, where $P_n$ is the $n \times n$ bit-reversal permutation matrix. Finally, $\boldsymbol{X}$ is transmitted through $n$ independent copies of $W$, as shown below:

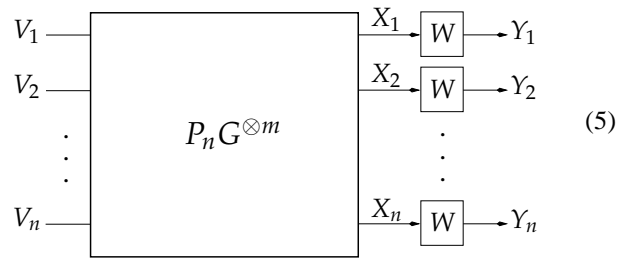

(5)

Arıkan [3] considered the $n$ channels "seen" by each of the $n$ individual bits $V_1, V_2, \ldots, V_n$ as they undergo the transformation in (5). Let us call them the *bit-channels* — for a precise definition of the notion of a bit-channel, see [3] and Section III. It is shown in [3] that as $m$ grows, the bit-channels start polarizing: they approach either a noiseless channel or a pure-noise channel. We will say that the former bit-channels are *good* while the latter are *bad* (again, see Section III for a rigorous definition). One of the key results of [3] is that the fraction of bit-channels that are good approaches the capacity of $W$ as $n \to \infty$.

Given the channel polarization phenomenon, the general idea of our construction is quite simple. We will transmit random bits over those bit-channels that are good for both Eve and Bob, information bits over those bit-channels that are good for Bob but bad for Eve, and zeros over those bit-channels that are bad for both Bob and Eve. In the rest of this paper, we make this idea precise and prove that it works.

In Section II, we briefly recap relevant results from the literature on wiretap channels, in order to obtain a simple expression for the secrecy capacity $\mathcal{C}_s$ in the case where $C_1$ and $C_2$ are symmetric DMCs and $C_2$ is degraded with respect to $C_1$. In Section III, we provide the necessary background on polar codes and establish a certain property of channel polarization that is crucial for our construction (Lemma 4). The construction itself, namely the proposed coding scheme, is presented in Section IV. In Section V, we prove that the proposed coding scheme satisfies the reliability and security conditions (1) and (2). We also show in Section V that the rate $k/n$ of our coding scheme approaches the secrecy capacity $\mathcal{C}_s$ as $n \to \infty$.

In Section VI, we consider the stronger notion of security (3). It was shown by Maurer-Wolf [28] that any coding scheme that satisfies the "weak" security condition (2) can be converted into a coding scheme that satisfies the stronger condition (3). This is accomplished using an ingenious information reconciliation and privacy amplification protocol [6]. Although, in principle, the rate overhead necessary for privacy amplification can be made arbitrarily small, this is unlikely to be the case in practice. In Section VI, we show how to modify the coding scheme of Section IV in order to guarantee strong security *directly*, without the need for privacy amplification. This is achieved by suitably modifying our definition of "bad" bit-channels, in a manner that differs from the generally accepted notions [3]. As a result, under the modified definition, a vanishing fraction of bit-channels could be good for Eve but bad for Bob, even when the wiretap channel is degraded with respect to the main channel. In this situation, the rate of the coding scheme of Section VI still approaches the secrecy capacity $\mathcal{C}_s$, but we cannot guarantee that the reliability condition (1) is satisfied, unless the main channel is noiseless. Nevertheless, we believe that, in practice, acceptably low probabilities of error could be achieved on the main channel (using a more elaborate decoding algorithm).

We conclude the paper in Section VII with a brief discussion of further results. In particular, we explain in Section VII that our construction generalizes straightforwardly to the case where the channels $C_1$ and $C_2$ are *not* symmetric, although in this case polar codes become less explicit and only the "symmetric secrecy capacity" can be achieved. A few open problems that stem from our results herein are also discussed in Section VII.

### C. Related Work

Following the publication of a preliminary version of this paper in [24] and [25], several related papers have appeared [1, 16, 20]. Most notably, the work of Hof and Shamai [16] on polar coding for wiretap channels is independent and contemporaneous to ours. While some of the main results in [16] and in this paper are similar, there are important differences that we would like to emphasize. One key difference is that Hof and Shamai [16] analyze their polar-coding scheme in detail, including recursive channel combining and splitting, whereas we treat polar codes essentially as a black box. We believe this makes our proof both shorter and clearer.

There are also significant differences between the results established in [1, 16, 20] and in this paper. In particular, it is shown in [1, 16] that polar coding achieves the entire *rate-equivocation region* (see [23] for a definition), whereas we are interested only in the extreme point of this region that corresponds to secrecy capacity. On the other hand, in several other respects, our results are stronger than those of [1, 16, 20]. First, the proof in [1, 16] is contingent on the assumption that the message $U$ is a priori uniform over $\{0, 1\}^k$, whereas we do not place any constraints on the a priori distribution of $U$. Assuming that messages are a priori uniform is common in information theory, but such assumptions are completely unacceptable in cryptography [5, 14]. Even more importantly, we show how polar coding should be used to provide *strong security*, whereas the work of [1, 16, 20] provides weak security only. Again, in cryptographic applications, conventional weak security is usually unacceptable.

## II. Secrecy Capacity

In this section, we first establish some relevant terminology. We then briefly recap the results of [11, 21] to provide a simple expression for the secrecy capacity $\mathcal{C}_s$ in the case where $C_1$ and $C_2$ are symmetric DMCs and $C_2$ is degraded with respect to $C_1$.

We will limit our consideration to finite-input and finite-output discrete memoryless channels throughout. Such a **channel** is a triple $\langle \mathscr{X}, \mathscr{Y}, W \rangle$, where $\mathscr{X}, \mathscr{Y}$ are finite sets and $W$ is an $|\mathscr{X}| \times |\mathscr{Y}|$ matrix with $W[x, y]$ being the probability of receiving $y \in \mathscr{Y}$ given that $x \in \mathscr{X}$ was sent. We will follow the convention of [3, 4, 18] and write $W(y|x)$ instead of $W[x, y]$.

A matrix $M$ is strongly symmetric if the rows of $M$ are permutations of each other and the columns of $M$ are permutations of each other. A channel $\langle \mathscr{X}, \mathscr{Y}, W \rangle$ is **strongly symmetric** if $W$ is a strongly symmetric matrix. Following [3, 4, 13, 18], we will say that $\langle \mathscr{X}, \mathscr{Y}, W \rangle$ is **symmetric** (often called *output-symmetric*) if the columns of $W$ can be partitioned into subsets such that each subset forms a strongly symmetric matrix. The **capacity** of a symmetric channel $\langle \mathscr{X}, \mathscr{Y}, W \rangle$ is given by

$$\mathcal{C}(W) \stackrel{\text{def}}{=} H(X) - H(X|Y) = \log_2 |\mathscr{X}| - H(X|Y) \quad (6)$$

where the random variable $X$ at the input to the channel is uniform over $\mathscr{X}$, and $Y$ is the corresponding random variable at the channel output (for a proof of this fact, see [13, p. 94]). An important example of a symmetric channel is the **binary symmetric channel** $\mathrm{BSC}(p) = \langle \{0, 1\}, \{0, 1\}, W \rangle$ with

$$W = \begin{bmatrix} 1-p & p \\ p & 1-p \end{bmatrix}$$

Given a channel $C_1 = \langle \mathscr{X}, \mathscr{Y}, W_1 \rangle$, we say that another channel $C_2 = \langle \mathscr{X}, \mathscr{Z}, W_2 \rangle$ is **degraded with respect to** $C_1$ if there exists a third channel $C_3 = \langle \mathscr{Y}, \mathscr{Z}, W_3 \rangle$ such that $C_2$ is the cascade of $C_1$ and $C_3$. Specifically, it is required that

$$W_2(z|x) = \sum_{y \in \mathscr{Y}} W_1(y|x) W_3(z|y) \quad (7)$$

for all $x \in \mathscr{X}$ and $z \in \mathscr{Z}$. Note that whenever $p_2 \geqslant p_1$, the channel $C_2 = \mathrm{BSC}(p_2)$ is degraded with respect to $C_1 = \mathrm{BSC}(p_1)$.

The *secrecy capacity* $\mathcal{C}_s$ of the wiretap-channel system in Figure 1 is defined as follows. First, assume that the message $U$ is uniformly random over $\{0, 1\}^k$. Then $\mathcal{C}_s$ is the supremum over all rates $R = k/n$ (in bits per channel use) such that there exist coding schemes of rate $R$ satisfying conditions (1) and (2). For the general case where $C_1$ and $C_2$ are arbitrary DMCs, comput-ing the secrecy capacity is a difficult problem. Let $X$ denote the single-letter input to $C_1$ and $C_2$, let $Y$ and $Z$ denote the corresponding single-letter outputs. The best known expression for the secrecy capacity $C_s$, given by Csiszár and Körner in [11], is

$$C_s = \max_U \Big( I(U;Y) - I(U;Z) \Big)$$

where the maximum is taken over all random variables $U$ such that $U \to X \to (Y,Z)$ is a Markov chain. The problem is that this maximization is often difficult to evaluate, and there is no simpler expression for the secrecy capacity even when $C_1$ and $C_2$ are both strongly symmetric, unless additional constraints are satisfied. See [23,39] for more details on this.

However, when $C_1 = \langle \mathcal{X}, \mathcal{Y}, W^* \rangle$ and $C_2 = \langle \mathcal{X}, \mathcal{Z}, W \rangle$ are symmetric *and* $C_2$ is degraded with respect to $C_1$, a simple expression for $C_s$ was given by Leung-Yan-Cheong in [21]. It is shown in [21, Theorem 4] that in this case

$$C_s = \mathcal{C}(W^*) - \mathcal{C}(W) = H(X|Z) - H(X|Y) \quad (8)$$

where $X$ is uniform over $\mathcal{X}$. In particular, if the main channel is BSC($p_1$) while the wiretap channel is BSC($p_2$), with $p_2 \geqslant p_1$, then the secrecy capacity is given by $h_2(p_2) - h_2(p_1)$, where $h_2(\cdot)$ is the binary entropy function.

## III. Polar Codes

This section provides a concise overview of the groundbreaking work of Arıkan [3] and others [4,18,19] on polar codes and channel polarization. We establish only those results that are essential for the coding schemes presented in this paper.

As in [15,19], we consider exclusively *binary-input symmetric memoryless* (BSM) discrete channels. Such a channel is a symmetric DMC, as defined in the previous section, with input alphabet $\mathcal{X} = \{0,1\}$. With a slight abuse of notation, we will often follow [3,4,19] and simply write $W$ to denote a BSM channel $\langle \{0,1\}, \mathcal{Y}, W \rangle$. The *Bhattacharyya parameter* of $W$ is

$$Z(W) \stackrel{\text{def}}{=} \sum_{y \in \mathcal{Y}} \sqrt{W(y|0) W(y|1)}$$

It can be shown that $Z(W)$ always takes values in $[0,1]$. Intuitively, channels with $Z(W) \leqslant \varepsilon$ are almost noiseless, while channels with $Z(W) \geqslant 1 - \varepsilon$ are almost pure-noise channels. This intuition is made precise in [3, Proposition 1].

Arıkan [3] introduces a number of channels that are associated with the transformation in (5). First, there is the channel $\langle \{0,1\}^n, \mathcal{Y}^n, W^n \rangle$ given by

$$W^n(\boldsymbol{y}|\boldsymbol{x}) \stackrel{\text{def}}{=} \prod_{i=1}^n W(y_i|x_i) \quad (9)$$

where $\boldsymbol{x} = (x_1, x_2, \ldots, x_n)$ and $\boldsymbol{y} = (y_1, y_2, \ldots, y_n)$. This is the channel that results from $n$ independent uses of the channel $W$. Next, for all $n = 2^m$, let us define the *Arıkan transform matrix* $G_n \stackrel{\text{def}}{=} P_n G^{\otimes m}$, where $G$ is the matrix in (4) and $P_n$ is the bit-reversal permutation matrix defined in [3, Section VII-B]. Arıkan then introduces the "combined" channel $\langle \{0,1\}^n, \mathcal{Y}^n, \widetilde{W} \rangle$ with transition probabilities given by

$$\widetilde{W}(\boldsymbol{y}|\boldsymbol{v}) \stackrel{\text{def}}{=} W^n(\boldsymbol{y} | \boldsymbol{v} G_n) = W^n(\boldsymbol{y} | \boldsymbol{v} P_n G^{\otimes m}) \quad (10)$$

This is the channel seen by the random vector $(V_1, V_2, \ldots, V_n)$ as it undergoes the transformation in (5). Arıkan [3] also defines the channel $\langle \{0,1\}, \mathcal{Y}^n \times \{0,1\}^{i-1}, W_i \rangle$ that is seen by the $i$-th bit $V_i$, for $i = 1, 2, \ldots, n$, as follows. Let $\boldsymbol{v}_i = (v_1, v_2, \ldots, v_i)$ denote a binary vector of length $i$, with the convention that $\boldsymbol{v}_0$ is the empty string and that $\{0,1\}^0 = \{\boldsymbol{v}_0\}$. Then

$$W_i(\boldsymbol{y}, \boldsymbol{v}_{i-1}|v_i) \stackrel{\text{def}}{=} \frac{1}{2^{n-1}} \sum_{\overline{\boldsymbol{v}} \in \{0,1\}^{n-i}} \widetilde{W}\Big(\boldsymbol{y} \,\Big|\, (\boldsymbol{v}_{i-1}, v_i, \overline{\boldsymbol{v}})\Big) \quad (11)$$

where $(\cdot,\cdot)$ denotes vector concatenation. It is easy to show (cf. Lemma 14) that $W_i(\boldsymbol{y}, \boldsymbol{v}_{i-1}|v_i)$ is indeed the probability of the event that $(Y_1, Y_2, \ldots, Y_n) = \boldsymbol{y}$ and $(V_1, V_2, \ldots, V_{i-1}) = \boldsymbol{v}_{i-1}$ given the event $V_i = v_i$, provided $\boldsymbol{V} = (V_1, V_2, \ldots, V_n)$ is a priori uniform over $\{0,1\}^n$. Consequently, if one considers a "hypothetical decoder" that attempts to estimate the $i$-th bit $V_i$ having observed $\boldsymbol{y}$ and $\boldsymbol{v}_{i-1}$, then $W_i$ is the effective channel seen by such decoder (again, provided $\boldsymbol{V}$ is a priori uniform). We will refer to $W_i$ as the *$i$-th bit-channel*, for $i = 1, 2, \ldots, n$.

Observe that the optimal decision rule for the hypothetical decoder of the foregoing paragraph is trivial: decide $\widehat{v}_i = 0$ if

$$W_i(\boldsymbol{y}, \boldsymbol{v}_{i-1}|0) \geqslant W_i(\boldsymbol{y}, \boldsymbol{v}_{i-1}|1) \quad (12)$$

and $\widehat{v}_i = 1$ otherwise. One can invoke this decision rule iteratively for all $i = 1, 2, \ldots, n$, while substituting the first $i-1$ decisions $(\widehat{v}_1, \widehat{v}_2, \ldots, \widehat{v}_{i-1})$ in place of the hypothetical observations $\boldsymbol{v}_{i-1}$. Up to a small modification described later, this is the *successive cancellation decoder* invented by Arıkan [3].

Following [4,18], let us partition the $n$ bit-channels into *good channels* and *bad channels* as follows. Let $[n] \stackrel{\text{def}}{=} \{1, 2, \ldots, n\}$ and let $\beta < 1/2$ be a fixed positive constant. Then the index sets of the good and bad channels are given by

$$\mathcal{G}_n(W, \beta) \stackrel{\text{def}}{=} \Big\{ i \in [n] : Z(W_i) < 2^{-n^\beta}/n \Big\} \quad (13)$$

$$\mathcal{B}_n(W, \beta) \stackrel{\text{def}}{=} \Big\{ i \in [n] : Z(W_i) \geqslant 2^{-n^\beta}/n \Big\} \quad (14)$$

One of the key results of [3,4] is that the fraction of the good channels approaches the channel capacity $\mathcal{C}(W)$, given by (6), as $n \to \infty$. We state this result precisely as follows.

**Theorem 1.** *For any BSM channel $W$ and any constant $\beta < 1/2$ we have*

$$\lim_{n \to \infty} \frac{|\mathcal{G}_n(W, \beta)|}{n} = \mathcal{C}(W)$$

Theorem 1 readily leads to a construction of capacity-achieving *polar codes*. The general idea is to transmit the information bits over the good bit-channels while fixing the input to the bad bit-channels to a priori known values, say zeros[1]. Formally, given a vector $\boldsymbol{v}$ of length $n$ and a set $\mathcal{A} \subseteq [n]$, let $\boldsymbol{v}_{\mathcal{A}}$ denote the projection of $\boldsymbol{v}$ on the coordinates in $\mathcal{A}$. Each subset $\mathcal{A}$ of $[n]$ of size $|\mathcal{A}| = k$ specifies a *polar code* $\mathbb{C}_n(\mathcal{A})$ of rate $k/n$. We define $\mathbb{C}_n(\mathcal{A})$ via its encoder map $\mathcal{E} : \{0,1\}^k \to \{0,1\}^n$. Given a message $\boldsymbol{u} \in \{0,1\}^k$, the encoder proceeds in two steps. First, the encoder constructs the vector $\boldsymbol{v} \in \{0,1\}^n$, by setting $\boldsymbol{v}_{\mathcal{A}} = \boldsymbol{u}$ and $\boldsymbol{v}_{\mathcal{A}^c} = \boldsymbol{0}$, where $\mathcal{A}^c$ is the complement of $\mathcal{A}$ in $[n]$ and $\boldsymbol{0}$ is the all-zero vector. Next, it outputs $\mathcal{E}(\boldsymbol{u}) = \boldsymbol{v} G_n$ as in (5). The decoder we will use for $\mathbb{C}_n(\mathcal{A})$ is the successive cancellation decoder of Arıkan [3]. This decoder works as already described in (12), with one straightforward modification: for $i \in \mathcal{A}^c$, the decision rule is simply $\widehat{v}_i = 0$.

---

[1] In the work of [3,4], which deals with symmetric as well as non-symmetric DMCs, it is important to allow an arbitrary choice of these "frozen" values. However, it is also shown in [3] that in the case of symmetric channels, any choice is as good as any other. Since we are concerned exclusively with symmetric channels in this paper, we will use zeros for notational convenience.

The key property of the encoder-decoder pair of the foregoing paragraph is summarized in the following theorem. This theorem is (the second part of) Proposition 2 of Arıkan [3].

**Theorem 2.** *Let $W$ be a BSM channel and let $\mathcal{A}$ be an arbitrary subset of $[n]$ of size $|\mathcal{A}| = k$. Suppose that a message $\boldsymbol{U}$ is chosen <u>uniformly at random</u> from $\{0,1\}^k$, encoded as a codeword of $\mathbb{C}_n(\mathcal{A})$, and transmitted over $W$. Then the probability that the channel output is not decoded to $\boldsymbol{U}$ under successive cancellation decoding satisfies*

$$\Pr\{\widehat{\boldsymbol{U}} \neq \boldsymbol{U}\} \leqslant \sum_{i \in \mathcal{A}} Z(W_i) \tag{15}$$

In this paper, we need a result that is somewhat stronger than Theorem 2, since we do *not* assume that messages are chosen uniformly at random from $\{0,1\}^k$. Fortunately, such a result can be readily established using the machinery that was already developed by Arıkan [3] for symmetric channels. Indeed, for symmetric channels, it is well known that the error probability is independent of the transmitted codeword; hence, the input distribution should not matter. The following proposition makes this observation precise. We include a proof, for completeness.

**Proposition 3.** *Let $W$ be a BSM channel and let $\mathcal{A}$ be an arbitrary subset of $[n]$ of size $|\mathcal{A}| = k$. Suppose that a message $\boldsymbol{U}$ is chosen <u>according to an arbitrary distribution</u> from $\{0,1\}^k$, encoded as a codeword of $\mathbb{C}_n(\mathcal{A})$, and transmitted over $W$. Then the probability $P_e$ that the channel output is not decoded to $\boldsymbol{U}$ under successive cancellation decoding satisfies*

$$P_e \leqslant \sum_{i \in \mathcal{A}} Z(W_i) \tag{16}$$

*Proof.* Following Arıkan [3, Section V], we consider the sample space $\Omega_n = \{0,1\}^n \times \mathcal{Y}^n$ with the probability measure

$$\Pr\{(\boldsymbol{v},\boldsymbol{y})\} \stackrel{\text{def}}{=} 2^{-n} \widetilde{W}(\boldsymbol{y}|\boldsymbol{v}) \tag{17}$$

for all $\boldsymbol{v} \in \{0,1\}^n$ and $\boldsymbol{y} \in \mathcal{Y}^n$. On this probability space, Arıkan [3] defines the event $\mathscr{E}$ of block error as the set of all pairs $(\boldsymbol{v},\boldsymbol{y})$ in $\Omega_n$ such that the channel output $\boldsymbol{y}$ is not decoded to $\boldsymbol{v}_\mathcal{A}$ under successive cancellation decoding. Let us further define, for all $\boldsymbol{w} \in \{0,1\}^n$, the event

$$\mathscr{V}_{\boldsymbol{w}} \stackrel{\text{def}}{=} \{(\boldsymbol{v},\boldsymbol{y}) \in \Omega_n \, : \, \boldsymbol{v} = \boldsymbol{w}\} \tag{18}$$

It is shown in [3, Proposition 2] that under the probability measure in (17) we have

$$\Pr\{\mathscr{E}\} \leqslant \sum_{i \in \mathcal{A}} Z(W_i) \tag{19}$$

It is furthermore shown in [3, Section VI-B] that, provided ties in the decision rule (12) are broken at random, the events $\mathscr{E}$ and $\mathscr{V}_{\boldsymbol{w}}$ are independent for all $\boldsymbol{w}$. In other words,

$$\Pr\{\mathscr{E} \,|\, \mathscr{V}_{\boldsymbol{w}}\} = \Pr\{\mathscr{E}\} \qquad \text{for all } \boldsymbol{w} \in \{0,1\}^n \tag{20}$$

Now consider the situation where the a priori distribution on the messages in $\{0,1\}^k$ is a delta-function. That is, a specific message $\boldsymbol{u} \in \{0,1\}^k$ is always chosen with probability 1, and the input to the transformation in (5) is the vector $\boldsymbol{w}$ with $\boldsymbol{w}_\mathcal{A} = \boldsymbol{u}$ and $\boldsymbol{w}_{\mathcal{A}^c} = \boldsymbol{0}$. Let $P_e(\boldsymbol{u})$ denote the probability that the successive cancellation decoder does not decode the corresponding channel output $\boldsymbol{y}$ to $\boldsymbol{w}_\mathcal{A}$. Then it follows from (10) that

$$P_e(\boldsymbol{u}) = \sum_{\boldsymbol{y} \in \mathcal{F}(\boldsymbol{w}_\mathcal{A})} \widetilde{W}(\boldsymbol{y}|\boldsymbol{w}) \tag{21}$$

where $\mathcal{F}(\boldsymbol{w}_\mathcal{A})$ is the set of channel outputs that are not decoded to $\boldsymbol{w}_\mathcal{A}$ by the successive cancellation decoder. Observe that

$$\mathscr{E} \cap \mathscr{V}_{\boldsymbol{w}} = \left\{(\boldsymbol{v},\boldsymbol{y}) \in \Omega_n \, : \, \boldsymbol{v} = \boldsymbol{w} \text{ and } \boldsymbol{y} \in \mathcal{F}(\boldsymbol{w}_\mathcal{A})\right\}$$

Therefore, the probability of the event $\mathscr{E} \cap \mathscr{V}_{\boldsymbol{w}}$, under the probability measure in (17), can be expressed as

$$\Pr\{\mathscr{E} \cap \mathscr{V}_{\boldsymbol{w}}\} = \sum_{\boldsymbol{y} \in \mathscr{E} \cap \mathscr{V}_{\boldsymbol{w}}} 2^{-n} \widetilde{W}(\boldsymbol{y}|\boldsymbol{v}) = 2^{-n} \sum_{\boldsymbol{y} \in \mathcal{F}(\boldsymbol{w}_\mathcal{A})} \widetilde{W}(\boldsymbol{y}|\boldsymbol{w}) \tag{22}$$

It can be readily seen from (18) that $\Pr\{\mathscr{V}_{\boldsymbol{w}}\} = 2^{-n}$ for all $\boldsymbol{w}$. Hence, it follows from (20) that

$$\Pr\{\mathscr{E} \cap \mathscr{V}_{\boldsymbol{w}}\} = \Pr\{\mathscr{E} \,|\, \mathscr{V}_{\boldsymbol{w}}\} \Pr\{\mathscr{V}_{\boldsymbol{w}}\} = 2^{-n} \Pr\{\mathscr{E}\} \tag{23}$$

Combining (21), (22), (23), we conclude that $P_e(\boldsymbol{u}) = \Pr\{\mathscr{E}\}$. Since this holds for all $\boldsymbol{u} \in \{0,1\}^k$, we have

$$P_e = \sum_{\boldsymbol{u} \in \{0,1\}^n} P_e(\boldsymbol{u}) \Pr\{\boldsymbol{U} = \boldsymbol{u}\} = \Pr\{\mathscr{E}\}$$

for *any* probability distribution $\Pr\{\boldsymbol{U} = \boldsymbol{u}\}$ on $\{0,1\}^k$. Hence, the proposition now follows from (19). ∎

In order to establish our main result in Section V, we also need to consider a slightly different encoding and decoding scenario. In this variation, the encoder $\mathcal{E}'$ is no longer deterministic. Rather, it has access to random bits and selects $\boldsymbol{V}_{\mathcal{A}^c}$ at random, according to some fixed (but otherwise arbitrary) probability distribution on $\{0,1\}^{n-k}$. In all other respects, $\mathcal{E}'$ is identical to the encoder $\mathcal{E}$ for $\mathbb{C}_n(\mathcal{A})$ described above. The specific realization $\boldsymbol{v}_{\mathcal{A}^c}$ of $\boldsymbol{V}_{\mathcal{A}^c}$ is revealed to the decoder by a genie. The decoder uses successive cancellation, with the suitable modification: for $i \in \mathcal{A}^c$, the value of $\widehat{v}_i$ is not set to zero, but rather to the corresponding coordinate in the realization $\boldsymbol{v}_{\mathcal{A}^c}$ (revealed by the genie). Let $P'_e$ denote the probability of block error in this scenario. It is shown in [3, Section VI] that Theorem 2 still applies in this case, namely

$$P'_e \leqslant \sum_{i \in \mathcal{A}} Z(W_i) \tag{24}$$

The coding scheme presented in the next section relies crucially on one more result from the literature on polar codes. The following lemma was proved by Korada in [18, Lemma 4.7].

**Lemma 4.** *Let $W$ and $W^*$ be BSM channels such that $W$ is degraded with respect to $W^*$. For $n = 2^m$, let $W_1, W_2, \ldots, W_n$ and $W_1^*, W_2^*, \ldots, W_n^*$ denote the $n$ corresponding bit-channels. Then $W_i$ is degraded with respect to $W_i^*$ for all $i = 1, 2, \ldots, n$, and therefore $C(W_i) \leqslant C(W_i^*)$ and $Z(W_i) \geqslant Z(W_i^*)$.*

It follows immediately from Lemma 4 and (13) that if $W$ is degraded with respect to $W^*$, then the set of good bit-channels for $W$ is a subset of the set of good bit-channels for $W^*$. More precisely, we have $\mathcal{G}_n(W, \beta) \subseteq \mathcal{G}_n(W^*, \beta)$ for all constants $\beta$.

## IV. THE CODING SCHEME

We consider a special case of the wiretap-channel system of Figure 1, wherein both the main channel $C_1 = \langle \{0,1\}, \mathcal{Y}, W^* \rangle$ and Eve's wiretap channel $C_2 = \langle \{0,1\}, \mathcal{Z}, W \rangle$ are symmetric DMCs, and $C_2$ is degraded with respect to $C_1$. The proposed coding scheme is illustrated informally below:

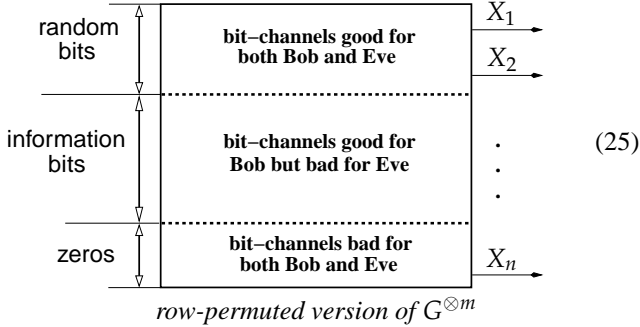

$$\text{(25)}$$

*row-permuted version of $G^{\otimes m}$*

The general idea is to transmit information *only* over those bit-channels that are bad for Eve, while flooding those bit channels that are good for Eve with random bits. Formally, we fix a positive constant $\beta < \frac{1}{2}$ and define three subsets of $[n]$ as follows:

$$\mathcal{R} \stackrel{\text{def}}{=} \mathcal{G}_n(W, \beta) \tag{26}$$

$$\mathcal{A} \stackrel{\text{def}}{=} \mathcal{G}_n(W^*, \beta) \setminus \mathcal{G}_n(W, \beta) \tag{27}$$

$$\mathcal{B} \stackrel{\text{def}}{=} \mathcal{B}_n(W^*, \beta) \tag{28}$$

Notice that the sets $\mathcal{R}, \mathcal{A}, \mathcal{B}$ are disjoint and $\mathcal{R} \cup \mathcal{A} \cup \mathcal{B} = [n]$. This is so since $\mathcal{G}_n(W^*, \beta)$ and $\mathcal{B}_n(W^*, \beta)$ are complements of each other by definition, and $\mathcal{G}_n(W, \beta) \subseteq \mathcal{G}_n(W^*, \beta)$ by Lemma 4. Let $|\mathcal{R}| = r$ and $|\mathcal{A}| = k$. We are now ready to describe the proposed encoding and decoding algorithms.

**Encoding Algorithm:** Formally, the encoder is a function $\mathcal{E}: \{0,1\}^k \times \{0,1\}^r \to \{0,1\}^n$. It accepts as input a message $u \in \{0,1\}^k$ and a vector $e \in \{0,1\}^r$. We make no assumptions about $u$ at this point, but we assume that $e$ is selected by Alice uniformly at random from $\{0,1\}^r$. The encoder first constructs the vector $v \in \{0,1\}^n$, by setting $v_\mathcal{R} = e$, $v_\mathcal{A} = u$, and $v_\mathcal{B} = 0$. The encoder then outputs $\mathcal{E}(u,e) := vG_n = vP_nG^{\otimes m}$ as in (5).

**Decoding Algorithm:** Formally, the decoder is a function $\mathcal{D}: \mathcal{Y}^n \to \{0,1\}^k$. It accepts as input a vector $y \in \mathcal{Y}^n$ at the output of the main channel $C_1 = \langle \{0,1\}, \mathcal{Y}, W^* \rangle$. It then invokes successive cancellation decoding for the polar code $\mathbb{C}_n(\mathcal{A} \cup \mathcal{R})$, used over $W^*$, to produce the vector $\widehat{v} \in \{0,1\}^n$. The decoder outputs $\mathcal{D}(y) := \widehat{v}_\mathcal{A}$.

We defer the proof that this coding scheme satisfies the reliability and security conditions (1), (2) to the next section. The rest of this section is devoted to two remarks about our construction.

**Remark.** We point out that our encoding algorithm can be regarded as a special case of the coset-coding method described in Section I-A. Recall that the coset-coding scheme is based upon an outer code $\mathbb{C}^*$ that provides error-correction on the main channel and an inner code $\mathbb{C} \subset \mathbb{C}^*$ that ensures security for the wiretap channel. In our encoding algorithm, the outer code is $\mathbb{C}^* = \mathbb{C}_n(\mathcal{A} \cup \mathcal{R})$ and the inner code is $\mathbb{C} = \mathbb{C}_n(\mathcal{R})$.   □

**Remark.** Given the channel polarization phenomenon, it's intuitively clear from (25) why the proposed coding scheme should work. The information bits $U = (U_1, U_2, \ldots, U_k)$ reach Bob via good (almost noiseless) bit-channels. Thus Bob should be able to reconstruct them with very high probability. On the other hand, these same bits pass through bad (almost pure-noise) bit-channels on their way to Eve. Thus Eve should not be able to deduce much information about $U$ from her observations $Z$, and $H(U|Z)$ should be close to $H(U)$.

However, this simple intuition is misleading, because it does not show how the random bits in (25) help keep Eve ignorant. It may appear that this randomness is not really needed. For example, what would happen if the vector $e$ that serves as the second input to our encoder function $\mathcal{E}(\cdot, \cdot)$ is not chosen at random from $\{0,1\}^r$ but rather set to an a priori fixed value? Since the channels are symmetric, any fixed value is as good as any other, so we may as well assume $e = 0$. This does not seem to affect the argument in the foregoing paragraph and, according to this argument, $H(U|Z)$ would still be close to $H(U)$.

In fact, this is not true. The reason is that channels seen by individual input bits as they undergo the transformation in (5) *depend on the distribution of other input bits*. Specifically, if $e = 0$ or $e$ is fixed, the resulting encoder will *not* be secure. This is an important point that we would like to establish rigorously. To do so, we first need the following simple lemma.

**Lemma 5.** *Let $\mathcal{S}$ be an arbitrary subset of $[n]$ of size $k$, and suppose that the polar code $\mathbb{C}_n(\mathcal{S})$ is used to communicate over a BSM channel $\langle \{0,1\}, \mathcal{Z}, W \rangle$. Further, assume that the message $U$ at the input to the encoder for $\mathbb{C}_n(\mathcal{S})$ is uniformly random over $\{0,1\}^k$, and let $Z$ denote the random vector at the channel output. Then $I(U; Z) \geqslant k\mathcal{C}(W)$.*

*Proof.* In fact, the lemma is true not only for $\mathbb{C}_n(\mathcal{S})$ but for any binary linear code of dimension $k$. The only property of the transform matrix $G_n$ that we need is that it is nonsingular.

Let $V$ be the random vector obtained by setting $V_\mathcal{S} = U$ and $V_{\mathcal{S}^c} = 0$. Then the codeword transmitted over the channel is

$$X = VG_n = UM \tag{29}$$

where $M$ is a $k \times n$ row submatrix of $G_n$. Since $G_n$ is nonsingular, $\text{rank}(M) = k$ and there exists a subset $\mathcal{T}$ of $[n]$ of size $k$ such that the corresponding $k$ columns of $M$ are linearly independent. This implies that there is a one-to-one correspondence between $U$ and $X_\mathcal{T}$. Hence, $I(U; Z) = I(X_\mathcal{T}; Z)$. Furthermore, since the random vector $U$ is uniform over $\{0,1\}^k$, so is the vector $X_\mathcal{T}$. Equivalently, its components $\{X_i : i \in \mathcal{T}\}$ are i.i.d. Ber($\frac{1}{2}$) random variables, and we can further conclude that

$$I(X_\mathcal{T}; Z) \geqslant I(X_\mathcal{T}; Z_\mathcal{T}) = \sum_{i \in \mathcal{T}} I(X_i; Z_i) = k\mathcal{C}(W)$$

where the last two equalities follow from the fact that the channel $\langle \{0,1\}, \mathcal{Z}, W \rangle$ is memoryless and symmetric.   ■

Now suppose that the input to our encoder function $\mathcal{E}(\cdot, \cdot)$ is a message chosen uniformly at random from $\{0,1\}^k$ along with $e = 0$. This is a special case of the situation considered in Lemma 5, with the set $\mathcal{S}$ given by (27). Hence $I(U; Z) \geqslant k\mathcal{C}(W)$, and security condition (2) cannot be satisfied: a significant fraction of message bits, at least $\mathcal{C}(W)$, is potentially exposed.   □

We note that the foregoing remark illustrates a *general result*. It is known that (2) cannot be satisfied unless the encoder makes use of at least $I(X;Z)$ random bits, where $I(X;Z)$ is the mutual information between the input and output of Eve's channel [26].

## V. Weak Security

In this section, we prove that the coding scheme of the previous section satisfies the reliability and security conditions (1) and (2) while its rate $k/n$ approaches the secrecy capacity.

The reliability of this coding scheme follows immediately from Proposition 3. Let $\widehat{V}$ denote the random vector at the output of the successive cancellation decoder for $\mathbb{C}_n(\mathcal{A} \cup \mathcal{R})$ invoked by our decoding algorithm, and note that the corresponding probability of block error is upper bounded by (16). Since $\widehat{U} = \widehat{V}_{\mathcal{A}}$ and $\mathcal{A} \cup \mathcal{R} = \mathcal{G}_n(W^*, \beta)$ by design, we see that

$$\Pr\{\widehat{U} \neq U\} \leqslant \sum_{i \in \mathcal{A} \cup \mathcal{R}} Z(W_i^*) \leqslant 2^{-n^\beta} \quad (30)$$

Since $n \geqslant k$, this clearly implies that $\lim_{k \to \infty} \Pr\{\widehat{U} \neq U\} = 0$ as required in (1), and the reliability condition is satisfied.

We now turn to the proof of security. For the remainder of this section, the sets $\mathcal{R}$, $\mathcal{A}$, $\mathcal{B}$ are given by (26)–(28), $U$ denotes Alice's message, $V$ denotes the intermediate vector constructed by our encoding algorithm (with $V_{\mathcal{A}} = U$, $V_{\mathcal{B}} = 0$, and $V_{\mathcal{R}}$ uniform over $\{0,1\}^r$), and $Z$ denotes Eve's observations. Also $|\mathcal{A}| = k$, $|\mathcal{R}| = r$, and $h_2(\cdot)$ is the binary entropy function.

**Lemma 6.**
$$H(V_{\mathcal{R}}|Z, V_{\mathcal{A}}) \leqslant h_2(2^{-n^\beta}) + r\, 2^{-n^\beta}$$

*Proof.* Suppose that in addition to her observations $Z$, a genie reveals to Eve the realization $v_{\mathcal{A}}$ of $V_{\mathcal{A}}$ and asks her to produce an estimate of $V_{\mathcal{R}}$. Eve also knows that $V_{\mathcal{B}} = 0$. Thus she knows all the bits of $V_{\mathcal{R}^c}$. Since $\mathcal{R} = \mathcal{G}_n(W, \beta)$, this is precisely the scenario considered in (24). Consequently, Eve can use successive cancellation decoding to deterministically compute an estimate $\widehat{V}_{\mathcal{R}} = f(Z, V_{\mathcal{A}})$ such that

$$\lambda \stackrel{\text{def}}{=} \Pr\{\widehat{V}_{\mathcal{R}} \neq V_{\mathcal{R}}\} \leqslant \sum_{i \in \mathcal{R}} Z(W_i) \leqslant 2^{-n^\beta} \quad (31)$$

We now invoke Fano's inequality [10, p. 38] to bound the conditional entropy $H(V_{\mathcal{R}}|Z, V_{\mathcal{A}})$ in terms of $\lambda$ as follows

$$H(V_{\mathcal{R}}|Z, V_{\mathcal{A}}) \leqslant h_2(\lambda) + r\lambda \quad (32)$$

where we have also used the fact that $V_{\mathcal{R}}$ takes values in the set $\{0,1\}^r$ of size $2^r$. The lemma now follows from (31), (32), and the fact that $h_2(\cdot)$ is increasing on the interval $[0, \tfrac{1}{2}]$. ∎

Let us define $\epsilon_n \stackrel{\text{def}}{=} \mathcal{C}(W) - |\mathcal{R}|/n$. Since $\mathcal{R} = \mathcal{G}_n(W, \beta)$, Theorem 1 implies that $\lim_{n \to \infty} \epsilon_n = 0$.

**Lemma 7.**
$$I(U;Z) \leqslant n\epsilon_n + h_2(2^{-n^\beta}) + (n-k)\,2^{-n^\beta} \quad (33)$$

*Proof.* The lemma is proved via a long sequence of simple equalities and inequalities, as follows:

$$I(U;Z) = I(V_{\mathcal{A}};Z) = I(V_{\mathcal{A} \cup \mathcal{B}};Z) \quad (34)$$
$$= I(V;Z) - I(V_{\mathcal{R}};Z|V_{\mathcal{A} \cup \mathcal{B}}) \quad (35)$$
$$= I(V;Z) - I(V_{\mathcal{R}};Z|V_{\mathcal{A}}) \quad (36)$$
$$= I(V;Z) - H(V_{\mathcal{R}}|V_{\mathcal{A}}) + H(V_{\mathcal{R}}|Z, V_{\mathcal{A}}) \quad (37)$$
$$= I(V;Z) - H(V_{\mathcal{R}}) + H(V_{\mathcal{R}}|Z, V_{\mathcal{A}}) \quad (38)$$
$$= I(V;Z) - r + H(V_{\mathcal{R}}|Z, V_{\mathcal{A}}) \quad (39)$$
$$\leqslant n\,\mathcal{C}(W) - r + H(V_{\mathcal{R}}|Z, V_{\mathcal{A}}) \quad (40)$$
$$= n\epsilon_n + H(V_{\mathcal{R}}|Z, V_{\mathcal{A}}) \quad (41)$$
$$\leqslant n\epsilon_n + h_2(2^{-n^\beta}) + r\,2^{-n^\beta} \quad (42)$$
$$\leqslant n\epsilon_n + h_2(2^{-n^\beta}) + (n-k)\,2^{-n^\beta} \quad (43)$$

The equalities in (34) hold since $V_{\mathcal{A}} = U$ and $V_{\mathcal{B}} = 0$. Now observe that $\mathcal{A} \cup \mathcal{B}$ and $\mathcal{R}$ are complements of each other in $[n]$. Hence, any distribution of $V$ can be thought of as a joint distribution of $V_{\mathcal{R}}$ and $V_{\mathcal{A} \cup \mathcal{B}}$. Given this observation, (35) follows from the chain rule for mutual information. The equality in (36) is trivial from $V_{\mathcal{B}} = 0$, while (37) is the definition of conditional mutual information. The equalities (38) and (39) hold since $V_{\mathcal{R}}$ and $V_{\mathcal{A}}$ are independent, and $V_{\mathcal{R}}$ is a priori uniform over $\{0,1\}^r$. Inequality (40) is immediate from the fact that $\mathcal{C}(W)$ is the capacity of $W$, while (41) follows from the definition of $\epsilon_n$. Finally, (42) follows from Lemma 6 and (43) is trivial. ∎

**Theorem 8.** *The encoding algorithm of the previous section satisfies the weak security condition* (2), *namely*

$$\lim_{k \to \infty} \frac{I(U;Z)}{k} = 0$$

*Proof.* This follows immediately from Lemma 7. Divide both sides of (33) by $n$ to get

$$\frac{I(U;Z)}{n} \leqslant \epsilon_n + \frac{h(2^{-n^\beta})}{n} + \frac{n-k}{n}2^{-n^\beta} \quad (44)$$

It is clear that the last two terms in (44) tend to zero as $n \to \infty$, and $\lim_{n \to \infty} \epsilon_n = 0$ by Theorem 1. Along with the obvious fact that $k = \Omega(n)$, this completes the proof of the theorem. ∎

Recall that for the wiretap-channel systems considered in this paper, the secrecy capacity is $\mathcal{C}_s = \mathcal{C}(W^*) - \mathcal{C}(W)$ (cf. Section II). Let $R_n = k/n$ denote the rate of our coding scheme.

**Theorem 9.**
$$\lim_{n \to \infty} R_n = \mathcal{C}(W^*) - \mathcal{C}(W)$$

*Proof.* Observe that

$$R_n = \frac{|\mathcal{A}|}{n} = \frac{|\mathcal{G}_n(W^*, \beta)|}{n} - \frac{|\mathcal{G}_n(W, \beta)|}{n} \quad (45)$$

where we have used the definition of $\mathcal{A}$ in (27) and the fact that $\mathcal{G}_n(W, \beta)$ is a subset of $\mathcal{G}_n(W^*, \beta)$ for all $\beta < \tfrac{1}{2}$. The theorem now follows from Theorem 1. ∎

Theorem 9 does not directly imply that our coding scheme achieves secrecy capacity, because the rate of communication from Alice to Bob, measured in information bits per channel use, could be much less than $k/n$ when $H(U) < k$. But this is true for *any* encoder that converts $k$ input bits to $n$ coded output bits. If the encoder accommodates an arbitrary distribution on its input $U$, it can achieve capacity *only* when $H(U) = k(1-o(1))$. If this necessary condition is satisfied, then our coding scheme does achieve secrecy capacity by Theorem 9.

## VI. STRONG SECURITY

This section shows how polar coding could be used to provide strong security whenever the main channel $C_1$ and the wiretap channel $C_2$ are symmetric binary-input DMCs, and $C_2$ is degraded with respect to $C_1$. Specifically, we describe a polar coding scheme that satisfies the strong security condition (3), while operating at a rate $k/n$ that approaches the secrecy capacity.

First, we will introduce a subtle but important change in the coding scheme of Section IV (see Section VI-B). In order to show that this change suffices to guarantee strong security, we need to replace the proof in the previous section by a more intricate argument (see Section VI-D). This argument relies crucially on the fact that a certain composite channel induced by our construction is symmetric (Section VI-C). Aided by a recent result of Hassani and Urbanke [15], we then prove that the rate of the proposed coding scheme approaches the secrecy capacity (Section VI-E). Unfortunately, we can show that the reliability condition (1) is satisfied only for the case where the main channel is noiseless. Nevertheless, we believe that, in practice, low probabilities of block error can be achieved also when the main channel is not noiseless, using a modification of Arıkan's successive cancellation decoder (Section VI-F).

### A. Analysis of the Weak-Security Coding Scheme

Henceforth, let us refer to the coding scheme introduced in Section IV and analyzed in the previous section as the ***weak-security coding scheme***. A natural question is whether this coding scheme does, in fact, provide strong security. Although we do not have a definitive answer to this question, we conjecture that it does not. As before, let

$$\epsilon_n \stackrel{\text{def}}{=} \mathcal{C}(W) - \frac{|\mathcal{R}|}{n} = \mathcal{C}(W) - \frac{|\mathcal{G}_n(W,\beta)|}{n} \quad (46)$$

with the sets $\mathcal{R}$ and $\mathcal{G}_n(W,\beta)$ defined as in (26) and (13). The following result provides some evidence for this conjecture.

**Proposition 10.** *Whenever the wiretap channel is a binary-input symmetric DMC and the main channel is noiseless, the weak-security coding scheme achieves strong security if and only if*

$$\lim_{n\to\infty} n\epsilon_n = 0 \quad (47)$$

*Proof.* The fact that (47) is sufficient for strong security is obvious from Lemma 7, since the last two terms in (33) tend to zero exponentially fast. In fact, it is clear that (47) is sufficient for strong security, whether the main channel is noiseless or not. We now show that this condition is also necessary, at least in the case where the main channel is noiseless. In this case, the set $\mathcal{B}$ in (28) is empty, and the vector $V$ at the input to the transformation (5) consists of $V_\mathcal{A} = U$ and $V_\mathcal{R}$, with $V_\mathcal{R}$ being uniform over $\{0,1\}^r$. Hence, if the message $U$ is a priori uniform over $\{0,1\}^k$, then $V$ is uniform over $\{0,1\}^n$. Let $X = VG_n$ as in (5). Since the Arıkan transform matrix $G_n$ is nonsingular, $X$ is also uniform over $\{0,1\}^n$. Consequently, we have

$$I(V;Z) = I(X;Z) = \sum_{i=1}^{n} I(X_i, Z_i) = n\mathcal{C}(W) \quad (48)$$

where the second equality follows by noting that $X_1, X_2, \ldots, X_n$ are i.i.d. $\text{Ber}(\tfrac{1}{2})$ random variables and $W$ is memoryless, while the last equality follows from the fact that $W$ is symmetric. This implies that the inequality (40) in the proof of Lemma 7 becomes an equality in this case. Therefore

$$I(U;Z) = n\epsilon_n + H(V_\mathcal{R}|Z, V_\mathcal{A}) \geqslant n\epsilon_n \quad (49)$$

It is now clear that $\lim_{n\to\infty} n\epsilon_n = 0$ is necessary for the mutual information $I(U;Z)$ to vanish asymptotically. ∎

Given a BSM channel $W$, is it true that $\lim_{n\to\infty} n\epsilon_n = 0$ for this channel? Unfortunately, the answer to this question is negative. It is known [33, 35] that for any discrete memoryless channel $W$ and *any* code of length $n$ and rate $R$ that achieves error-probability $P_e$ on $W$, we have

$$\mathcal{C}(W) - R \geqslant \frac{\text{const}(P_e, W)}{\sqrt{n}} - O\left(\frac{\log n}{n}\right)$$

where the constant (which is given explicitly in [33]) depends on $W$ and $P_e$, but not on $n$. This implies that $n\epsilon_n = \Omega(\sqrt{n})$, and the weak-security coding scheme does *not* provide strong security. Consequently, in order to provide strong security, the polar coding scheme of Section IV has to be modified.

### B. Strong-Security Coding Scheme

Intuitively, the main reason that the coding scheme of Section IV fails to provide strong security is this: the bit-channels that are deemed bad for Eve are not bad enough. Indeed, according to the definition of $\mathcal{B}_n(W,\beta)$ in (14), a bit-channel $W_i$ is considered bad for Eve whenever $Z(W_i) \geqslant 2^{-n^\beta}/n$. For example, if $n = 2^{10}$ and $\beta = 0.499$, a bit-channel may be declared bad for Eve even when its capacity is greater than $1 - 10^{-9}$ while Eve's probability of error on this channel is less than $2 \cdot 10^{-10}$. It is obvious that such a bit-channel does not prevent Eve from deducing the information at its input with high probability.

The problem is that the generally accepted definitions of good and bad bit-channels — for example, (13) and (14) — are motivated by *Bob's point of view*. First, a criterion for "goodness" is established, motivated by the probability of error on Bob's side, then the bit-channels that do not satisfy this criterion are deemed bad. In order to achieve strong security, we will re-define things from *Eve's point of view*. First, we introduce a strong criterion for "badness," and then make sure that random bits are sent over all the bit-channels that do not satisfy this criterion. Specifically, given a BSM channel $W$ and a positive $\delta < 1$, we define the index set of $\delta$-***poor bit-channels*** as follows:

$$\mathcal{P}_n(W,\delta) \stackrel{\text{def}}{=} \left\{ i \in [n] : \mathcal{C}(W_i) \leqslant \delta \right\} \quad (50)$$

Further, we leave the definition of the good bit-channels in (13) unchanged, but re-define the sets $\mathcal{R}$, $\mathcal{A}$, and $\mathcal{B}$ as follows:

$$\mathcal{R} \stackrel{\text{def}}{=} [n] \setminus \mathcal{P}_n(W,\delta_n) \quad (51)$$

$$\mathcal{A} \stackrel{\text{def}}{=} \mathcal{P}_n(W,\delta_n) \cap \mathcal{G}_n(W^*,\beta) \quad (52)$$

$$\mathcal{B} \stackrel{\text{def}}{=} \mathcal{P}_n(W,\delta_n) \setminus \mathcal{G}_n(W^*,\beta) \quad (53)$$

where, for the time being, $\delta_n$ is an arbitrary function from the positive integers to the interval $(0,1)$. We will specify this function precisely later in this section (cf. Theorem 17 and Proposi-tion 20). In the meantime, notice that the sets $\mathcal{R}$, $\mathcal{A}$, and $\mathcal{B}$, as defined in (51)–(53), are still disjoint and $\mathcal{R} \cup \mathcal{A} \cup \mathcal{B} = [n]$.

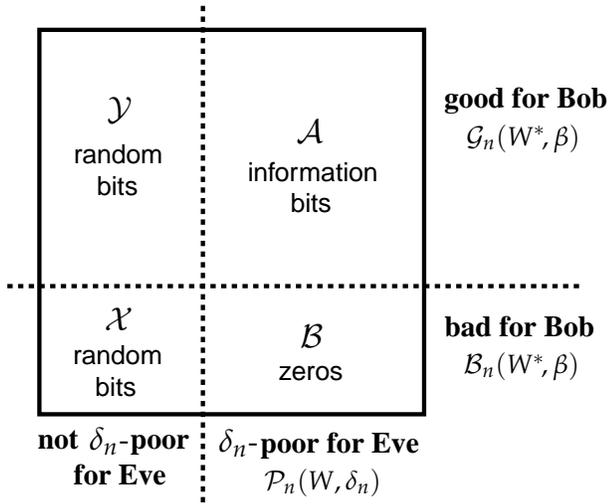

**Figure 2.** *Informal sketch of the strong-security coding scheme*

The *strong-security coding scheme* of (51)–(53) is illustrated schematically in Figure 2, wherein the bold square represents the set of $n$ bit-channels. This set is partitioned two ways: into bit-channels that are good for Bob and bad for Bob (as before), and into bit-channels that are $\delta_n$-poor and not $\delta_n$-poor for Eve. The sets $\mathcal{X}$ and $\mathcal{Y}$, with $\mathcal{X} \cup \mathcal{Y} = \mathcal{R}$, will be discussed later in this section (see (94), (95) and Section VI-F).

With the new definitions of the sets $\mathcal{R}$, $\mathcal{A}$, $\mathcal{B}$ in (51)–(53), the encoding and decoding algorithms for the strong-security coding scheme are exactly those given in Section IV for the weak-security coding scheme (although we will modify the decoding algorithm somewhat in Section VI-F).

### C. The Induced Channel Is Symmetric

We prove in the next subsection that the strong-security coding scheme indeed provides strong security. In order to do so, we first introduce and study a certain composite channel induced by our construction. Informally, this channel describes the transformation in (5) in the case where some $r$ of the bits $V_1, V_2, \ldots, V_n$ are set independently and uniformly at random, while the remaining $n - r$ bits serve as the input to the channel. This situation is depicted in Figure 3. Formally, the *induced channel* $\mathcal{Q}_n(W, \mathcal{R})$ is specified in terms of an arbitrary BSM channel $W$ with output alphabet $\mathscr{Z}$, and a subset $\mathcal{R}$ of $[n]$ of size $|\mathcal{R}| = r$. The input alphabet of $\mathcal{Q}_n(W, \mathcal{R})$ is $\{0,1\}^{n-r}$ and its output alphabet is $\mathscr{Z}^n$. To describe the transition probabilities of $\mathcal{Q}_n(W, \mathcal{R})$, let us introduce the following notation. Henceforth, given a vector $x \in \{0,1\}^{n-r}$ and a vector $e \in \{0,1\}^r$, let $(x; e)$ denote the vector $v \in \{0,1\}^n$ with $v_\mathcal{R} = e$ and $v_{\mathcal{R}^c} = x$. With this, referring to the definition of $W^n$ in (9), the $2^{n-r} \times |\mathscr{Z}|^n$ transition-probability matrix $Q$ of $\mathcal{Q}_n(W, \mathcal{R})$ is given by

$$Q(z|x) \stackrel{\text{def}}{=} \frac{1}{2^r} \sum_{e \in \{0,1\}^r} W^n(z \,|\, (x;e)\, G_n) \qquad (54)$$

for all $x \in \{0,1\}^{n-r}$ and all $z \in \mathscr{Z}^n$. It can be readily seen that if $V = (V_1, V_2, \ldots, V_{n-r})$ is the random vector at the input to the channel in Figure 3 and $Z = (Z_1, Z_2, \ldots, Z_n)$ is the random vector at the channel output, then indeed

$$Q(z|x) = \Pr\{Z = z \,|\, V = x\} \qquad (55)$$

Our main goal in this subsection is to prove that the induced channel $\mathcal{Q}_n(W, \mathcal{R})$ in Figure 3 is symmetric.

In order to do so, we start with a general result that uses elementary tools from group theory to provide a sufficient condition for a given channel to be symmetric. Recall that a *group action* of an abelian group $A$ on a set $\mathscr{Y}$ is a function from $A \times \mathscr{Y}$ to $\mathscr{Y}$, denoted $(a, y) \mapsto a.y$, with the following properties:

**P1.** $0.y = y$ for all $y \in \mathscr{Y}$, where 0 is the identity of $A$;
**P2.** $(a+b).y = a.(b.y)$ for all $a, b \in A$ and all $y \in \mathscr{Y}$, where $+$ denotes the group operation.

The orbit of $y \in \mathscr{Y}$ is the set of all points of $\mathscr{Y}$ to which $y$ can be moved by the elements of $A$. Explicitly, the *orbit of $y$* is

$$\mathcal{O}(y) \stackrel{\text{def}}{=} \{a.y \,:\, a \in A\} \qquad (56)$$

It is well known that orbits of points in $\mathscr{Y}$ form a partition of $\mathscr{Y}$ into equivalence classes (under the equivalence relation $y_1 \sim y_2$ iff there exists an $a \in A$ with $y_2 = a.y_1$).

**Theorem 11.** *Let $\langle \mathscr{X}, \mathscr{Y}, W \rangle$ be a DMC, and suppose that $\mathscr{X}$ is an abelian group under the binary operation $+$. Further, suppose that there exists a group action $.$ of $\mathscr{X}$ on $\mathscr{Y}$ such that*

$$W(y|a+x) = W(a.y|x) \qquad (57)$$

*for all $a, x \in \mathscr{X}$ and all $y \in \mathscr{Y}$. Then the DMC $\langle \mathscr{X}, \mathscr{Y}, W \rangle$ is necessarily a symmetric channel.*

*Proof.* We partition the set $\mathscr{Y}$ into orbits formed by the group action $.$ of $\mathscr{X}$. Let $\mathcal{O}$ be an orbit, and let $M$ be the $|\mathscr{X}| \times |\mathcal{O}|$ column submatrix of $W$ consisting of those columns that are indexed by the elements of $\mathcal{O}$. It would suffice to prove that $M$ is a strongly symmetric matrix, regardless of the choice of $\mathcal{O}$.

Consider two arbitrary rows of $M$ indexed, say, by the elements $x_1 \in \mathscr{X}$ and $x_2 \in \mathscr{X}$. Set $a = x_2 + (-x_1)$, where $-x_1$ is the inverse of $x_1$ in the group $\mathscr{X}$. Then we have $x_2 = a + x_1$. Therefore, (57) implies that

$$W(y|x_2) = W(a.y|x_1) \qquad \text{for all } y \in \mathscr{Y} \qquad (58)$$

It is easy to see from (56) and property P2 that the map $y \mapsto a.y$ is bijective on $\mathcal{O}$. Together with (58), this shows that the rows of $M$, indexed by $x_1$ and $x_2$, are permutations of each other.

Now consider two arbitrary columns of $M$ indexed by the elements $y_1 \in \mathcal{O}$ and $y_2 \in \mathcal{O}$. Then, by the definition of an orbit, there exists an $a \in \mathscr{X}$ such that $y_2 = a.y_1$, and (57) implies that

$$W(y_2|x) = W(y_1|a+x) \qquad \text{for all } x \in \mathscr{X} \qquad (59)$$

It is clear that the map $x \mapsto a+x$ is bijective on $\mathscr{X}$. Thus (59) shows that the columns of $M$ are permutations of each other. ∎

Notice that the input alphabet $\{0,1\}^{n-r}$ of $\mathcal{Q}_n(W, \mathcal{R})$ is an abelian group, with the group operation $+$ being the componentwise modulo 2 addition of vectors in $\mathbb{F}_2^{n-r}$. Consequently, in order to prove that the channel $\mathcal{Q}_n(W, \mathcal{R})$ is symmetric, it would suffice to construct a group action of $\{0,1\}^{n-r}$ on $\mathscr{Z}^n$ that satisfies (57). To do so, we will use the fact that $W$ itself is symmetric. As noted by Arıkan [3], a binary-input channel $W$ with output alphabet $\mathscr{Z}$ is symmetric if and only if there exists a permutation $\pi_1$ on $\mathscr{Z}$ such that

$$\pi_1 = \pi_1^{-1} \qquad (\pi_1 \text{ is an involution}) \tag{60}$$

$$W(z|0) = W(\pi_1(z)|1) \qquad \text{for all } z \in \mathscr{Z} \tag{61}$$

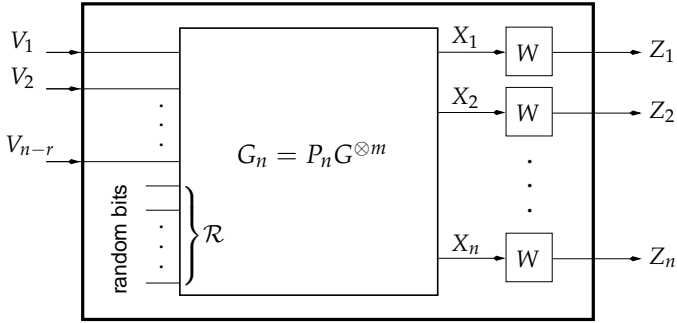

**Figure 3.** *Block diagram of the induced channel $\mathcal{Q}_n(W, \mathcal{R})$*

Let $\pi_0$ be the identity permutation on $\mathscr{Z}$. Following Arıkan [3], let us define a group action of the additive group of $\mathbb{F}_2 = \{0, 1\}$ on the set $\mathscr{Z}$ as follows: $x.z = \pi_x(z)$ for all $x \in \mathbb{F}_2$ and $z \in \mathscr{Z}$. It is trivial to verify that $x.z$ has the required group-action properties P1 and P2, and that for all $a, x \in \mathbb{F}_2$ and $z \in \mathscr{Z}$, we have

$$W(z|a + x) = W(a.z|x) \tag{62}$$

As in [3], we can extend this function componentwise to a group action of the additive group of $\mathbb{F}_2^n$ on the set $\mathscr{Z}^n$ as follows:

$$x.z \stackrel{\text{def}}{=} (x_1.z_1, x_2.z_2, \ldots, x_n.z_n) \tag{63}$$

The following lemma was proved by Arıkan in [3, Propositions 12 and 13]. We provide a simple proof herein, for completeness.

**Lemma 12.** *Let $G_n = P_n G^{\otimes m}$ be the Arıkan transform matrix. Then for all $a, x \in \mathbb{F}_2^n$ and all $z \in \mathscr{Z}^n$, we have*

$$W^n(z|(a + x)G_n) = W^n(aG_n.z|xG_n) \tag{64}$$

*Proof.* In fact, the lemma is true for an arbitrary $n \times n$ binary matrix (in other words, any linear transformation from $\mathbb{F}_2^n$ to itself). First, let us show that

$$W^n(z|b + c) = W^n(b.z|c) \tag{65}$$

for all $b, c \in \mathbb{F}_2^n$ and all $z \in \mathscr{Z}^n$. This follows directly from the definition of $W^n$. Indeed, expanding both sides of (65) as

$$\prod_{i=1}^n W(z_i|b_i + c_i) = \prod_{i=1}^n W(b_i.z_i|c_i)$$

we conclude that (65) is implied by (62). Let us now set $b = aG_n$ and $c = xG_n$. Then the lemma follows from (65) along with the fact that multiplication by a matrix is a linear operation, that is $(a + x)G_n = aG_n + xG_n = b + c$. ∎

We now depart from Arıkan [3], and introduce a group action $\circ$ of the additive group of $\mathbb{F}_2^{n-r}$ on $\mathscr{Z}^n$, defined as follows. Recall that $(x, e)$ denotes the vector $v \in \{0, 1\}^n$ with $v_{\mathcal{R}} = e$ and $v_{\mathcal{R}^c} = x$. With this, we define for all $x \in \mathbb{F}_2^{n-r}$ and all $z \in \mathscr{Z}^n$

$$x \circ z \stackrel{\text{def}}{=} (x; 0)G_n.z \tag{66}$$

where the group action . on the right-hand side is the one defined in (63). Again, it is easy to verify that (66) satisfies P1 and P2. Our main result in this subsection is the following proposition.

**Proposition 13.** *The induced channel $\mathcal{Q}_n(W, \mathcal{R})$ is symmetric.*

*Proof.* In light of Theorem 11, it would suffice to prove that the transition-probability matrix $Q$ defined in (54) satisfies

$$Q(z|a + x) = Q(a \circ z|x)$$

for all $a, x \in \mathbb{F}_2^{n-r}$ and $z \in \mathscr{Z}^n$. Expanding the vector $(a + x; e)$ as $(a; 0) + (x; e)$, and substituting in (54), we obtain

$$Q(z|a + x) = \frac{1}{2^r} \sum_{e \in \{0,1\}^r} W^n\big(z \,\big|\, ((a; 0) + (x; e))G_n\big) \tag{67}$$

$$= \frac{1}{2^r} \sum_{e \in \{0,1\}^r} W^n\big((a; 0)G_n.z \,\big|\, (x; e)G_n\big) \tag{68}$$

$$= \frac{1}{2^r} \sum_{e \in \{0,1\}^r} W^n\big(a \circ z \,\big|\, (x; e)G_n\big) \tag{69}$$

where (68) follows from Lemma 12 and (69) follows from (66). But (69) is precisely $Q(a \circ z | x)$ by (54), and we are done. ∎

### D. Proof of Strong Security

Let us begin with a simple lemma that relates the capacity of the bit-channels in (11) to the transformation in (5). Although this lemma is well-known, we include a proof for completeness.

**Lemma 14.** *Let $W$ be an arbitrary BSM channel. Suppose the vector $V = (V_1, V_2, \ldots, V_n)$ at the input to the transformation in (5) is uniform over $\{0, 1\}^n$, and let $Z = (Z_1, Z_2, \ldots, Z_n)$ be the random vector at the output of the transformation. Further, let $W_1, W_2, \ldots, W_n$ be the corresponding bit-channels, defined in (11). Then for all $i \in [n]$, the capacity of $W_i$ is given by*

$$\mathcal{C}(W_i) = I(V_i; Z, V_1, V_2, \ldots, V_{i-1})$$

*Proof.* Let $V_i$ denote the random vector $(V_1, V_2, \ldots, V_i)$ for all $i \in [n]$, as before. It is shown in [3] that if $W$ is symmetric, then so is $W_i$ for all $i \in [n]$. Hence, the capacity of $W_i$ is the mutual information between its input and output when the input is uniform over $\{0, 1\}$. Thus it would suffice to show that

$$W_i(z, v | x) = \Pr\{Z = z, V_{i-1} = v \,|\, V_i = x\} \tag{70}$$

for all $z \in \mathscr{Z}^n, v \in \{0, 1\}^{i-1}$, and $x \in \{0, 1\}$. Since $V_i$ is uniform we can re-write the right-hand side of (70) as follows:

$$\frac{\Pr\{Z = z, V_{i-1} = v, V_i = x\}}{\Pr\{V_i = x\}} = 2\Pr\{Z = z, V_i = (v, x)\}$$

Observe that the event $\{V_i = (v, x)\}$ is the union of $2^{n-i}$ disjoint events $\{V = (v, x, \overline{v})\}$, as $\overline{v}$ ranges over $\{0, 1\}^{n-i}$ (or $\overline{v}$ is the empty string, if $i = n$). Since $V$ is uniform over $\{0,1\}^n$, the probability of each such event is $2^{-n}$. Consequently

$$2 \Pr\{Z = z, V_i = (v,x)\} =$$
$$= 2 \sum_{\overline{v} \in \{0,1\}^{n-i}} \Pr\{Z = z, V = (v,x,\overline{v})\} \quad (71)$$
$$= \frac{1}{2^{n-1}} \sum_{\overline{v} \in \{0,1\}^{n-i}} \Pr\{Z = z \mid V = (v,x,\overline{v})\} \quad (72)$$

Since $Z$ and $V$ are, respectively, the output and the input to the transformation in (5), for all $w \in \{0,1\}^n$ we have

$$\Pr\{Z = z \mid V = w\} = W^n(z \mid w P_n G^{\otimes m}) = \widetilde{W}(z \mid w)$$

by the definition of the "combined" channel $\widetilde{W}$ in (10). Together with (72) and the definition of $W_i$ in (11), this shows that the right-hand side of (70) is indeed equal to $W_i(z,v \mid x)$. ∎

The next lemma combines Proposition 13 with Lemma 14 to upper-bound the capacity of the induced channel $\mathcal{Q}_n(W, \mathcal{R})$.

**Lemma 15.** *Let $W$ be an arbitrary BSM channel. For $n = 2^m$, let $W_1, W_2, \ldots, W_n$ denote the corresponding bit-channels. Then for all $\mathcal{R} \subset [n]$, the capacity of the induced channel $\mathcal{Q}_n(W, \mathcal{R})$, defined in (54), is upper-bounded as follows:*

$$\mathcal{C}(\mathcal{Q}_n(W, \mathcal{R})) \leqslant \sum_{i \in \mathcal{R}^c} \mathcal{C}(W_i) \quad (73)$$

*Proof.* Consider again the transformation in (5), with the input vector $V = (V_1, V_2, \ldots, V_n)$ being uniform over $\{0,1\}^n$, and let $Z = (Z_1, Z_2, \ldots, Z_n)$ denote the output of the transformation, as in Lemma 14. Then

$$\mathcal{C}(\mathcal{Q}_n(W, \mathcal{R})) = I(V_{\mathcal{R}^c}; Z) \quad (74)$$

This is so because $\mathcal{Q}_n(W, \mathcal{R})$ is symmetric by Proposition 13, and the capacity of a symmetric DMC is given by the mutual information between its input and output, *under uniform input distribution* [13, Theorem 4.5.2]. Next, write $\mathcal{R}^c = [n] \setminus \mathcal{R}$ as

$$\mathcal{R}^c = \{i_1, i_2, \ldots, i_{n-r}\}$$

where $r = |\mathcal{R}|$, and assume w.l.o.g. that $i_1 < i_2 < \cdots < i_{n-r}$. The lemma can be now proved via a sequence of simple equalities and inequalities, as follows:

$$I(V_{\mathcal{R}^c}; Z) = I(V_{i_1}, V_{i_1}, \ldots, V_{i_{n-r}}; Z) \quad (75)$$
$$= \sum_{j=1}^{n-r} I(V_{i_j}; Z \mid V_{i_1}, V_{i_2}, \ldots, V_{i_{j-1}}) \quad (76)$$
$$= \sum_{j=1}^{n-r} I(V_{i_j}; Z, V_{i_1}, V_{i_2}, \ldots, V_{i_{j-1}}) \quad (77)$$
$$\leqslant \sum_{j=1}^{n-r} I(V_{i_j}; Z, V_1, V_2, \ldots, V_{i_{j-1}}) \quad (78)$$

The equality (76) is the chain rule for mutual information. The equality (77) follows from the fact that $I(X; Z \mid Y) = I(X; Z, Y)$ for all random variables $X, Y, Z$ such that $X$ and $Y$ are independent. To establish (78), we adjoin to the set of random variables $\{V_{i_1}, V_{i_2}, \ldots, V_{i_{j-1}}\}$ its complement in the set $\{V_1, V_2, \ldots, V_{i_j}\}$.

Clearly, this cannot decrease the mutual information. Lastly, we observe that the summation in (78) is equal to the summation on the right-hand side of (73) by Lemma 14. ∎

With Lemma 15 in hand, we are finally ready to establish our main result in this section.

**Proposition 16.** *Let $U$ be the message at the input to the encoder for the strong-security coding scheme. Then, regardless of the a priori distribution of $U$, we have*

$$I(U; Z) \leqslant \delta_n |\mathcal{P}_n(W, \delta_n)| \quad (79)$$

*where $Z$ is the output of the wiretap channel, and $\mathcal{P}_n(W, \delta_n)$ is the index set of $\delta_n$-poor bit-channels, as defined in (50).*

*Proof.* Recall that our encoder first constructs the vector $V$ with $V_{\mathcal{A}} = U$, $V_{\mathcal{B}} = 0$, and $V_{\mathcal{R}}$ uniform over $\{0,1\}^r$. Hence

$$I(U; Z) = I(V_{\mathcal{A}}; Z) = I(V_{\mathcal{A} \cup \mathcal{B}}; Z)$$

Since the sets $\mathcal{A}, \mathcal{B}, \mathcal{R}$ partition $[n]$, the vector $V_{\mathcal{A} \cup \mathcal{B}}$ can be regarded as the input to the induced channel $\mathcal{Q}_n(W, \mathcal{R})$. Moreover, since $\mathcal{R}^c = \mathcal{P}_n(W, \delta_n)$ by (51), Lemma 15 implies that

$$I(V_{\mathcal{A} \cup \mathcal{B}}; Z) \leqslant \mathcal{C}(\mathcal{Q}_n(W, \mathcal{R})) \leqslant \sum_{i \in \mathcal{P}_n(W, \delta_n)} \mathcal{C}(W_i)$$

The proposition now follows by observing that $\mathcal{C}(W_i) \leqslant \delta_n$ for all $i \in \mathcal{P}_n(W, \delta_n)$, by the definition of $\mathcal{P}_n(W, \delta_n)$ in (50). ∎

Note that we are still free to specify the function $\delta_n$ in (51) and (79). This means that the security of our coding scheme is *tunable*. Let us henceforth refer to $\delta_n$ as the **security function**; this function is a *design parameter* in our scheme. Proposition 16 implies that choosing different settings for the security function guarantees different levels of security.

**Theorem 17.** *For any security function such that $\delta_n = o(1/n)$, the strong-security coding scheme guarantees strong security.*

*Proof.* Follows from Proposition 16, along with the definition of strong security in (3) and the fact that $|\mathcal{P}_n(W, \delta_n)| \leqslant n$. ∎

In fact, we shall see in the next subsection (cf. Theorem 21) that we can achieve the secrecy capacity, while setting the security function to be as small as $\delta_n = 2^{-n^\beta}$ for any positive constant $\beta < \frac{1}{2}$. In this case, our coding scheme guarantees that the mutual information between the message $U$ and Eve's observations $Z$ scales roughly as

$$I(U; Z) = o\left(2^{-\sqrt{n}/n^\varepsilon}\right) \quad (80)$$

for any $\varepsilon > 0$. Note that this holds regardless of the a priori distribution of the message.

*E. Rate of the Strong-Security Coding Scheme*

Let $R_n = k/n = |\mathcal{A}|/n$ denote the rate of the strong-security coding scheme, where $\mathcal{A}$ is the set defined in (52). Note that

$$\mathcal{A} = \mathcal{P}_n(W, \delta_n) \setminus \mathcal{B}_n(W^*, \beta)$$

since the sets $\mathcal{G}_n(W^*, \beta)$ and $\mathcal{B}_n(W^*, \beta)$ of good and bad channels in (13), (14) are complements of each other. It follows that

$$R_n \geqslant \frac{|\mathcal{P}_n(W, \delta_n)|}{n} - \frac{|\mathcal{B}_n(W^*, \beta)|}{n} \quad (81)$$

The asymptotic behavior of the fraction $|\mathcal{B}_n(W^*,\beta)|/n$ is given by Theorem 1. Therefore, in order to prove that the rate of the strong-security coding scheme approaches secrecy capacity, it remains to analyze the asymptotic behavior of $|\mathcal{P}_n(W,\delta_n)|/n$.

In this regard, a recent result of Hassani and Urbanke [15] will be useful. To describe this result, we need to introduce some notation. Given a BSM channel $W$ and a positive $\gamma < 1$, let

$$\mathcal{P}'_n(W,\gamma) \stackrel{\text{def}}{=} \left\{ i \in [n] \,:\, Z(W_i) \geqslant 1 - \gamma \right\} \quad (82)$$

This is similar to the definition of the set $\mathcal{P}_n(W,\delta)$ in (50), except that (82) uses the Bhattacharyya parameters $Z(W_i)$ instead of the channel capacities $\mathcal{C}(W_i)$. Given a positive integer $m$ and a real number $\xi$ in the open interval $(0,1)$, let $a = a(m,\xi)$ denote the unique positive integer such that

$$\sum_{i=a}^{m}\binom{m}{i} \leqslant \xi 2^m < \sum_{i=a-1}^{m}\binom{m}{i} \quad (83)$$

and define $\alpha(m,\xi) \stackrel{\text{def}}{=} a(m,\xi)/m$. Further, we say that a function $f(m)$ from the positive integers to the interval $(0,1)$ is *in the intersection of $o(1/\sqrt{m})$ and $\omega(1/m)$* if

$$\lim_{m\to\infty}\bigl(\sqrt{m}f(m)\bigr) = 0 \quad \text{and} \quad \lim_{m\to\infty}\bigl(mf(m)\bigr) = \infty$$

The following is (the second part of) Theorem 3 of Hassani and Urbanke [15]. Although this result is more general than what we need, we state it below exactly as in [15, Theorem 3].

**Theorem 18.** *Let $W$ be a BSM channel, and let $\xi < 1$ be a positive constant. Fix an arbitrary function $f(m)$ in the intersection of $o(1/\sqrt{m})$ and $\omega(1/m)$, and for all $n = 2^m$ define*

$$\gamma_n \stackrel{\text{def}}{=} 2^{-n^{\alpha(m,\xi)(1+f(m))}} \quad (84)$$

*where $\alpha(m,\xi)$ is the function defined in (83). Then the asymptotic behavior of the fraction $|\mathcal{P}'_n(W,\gamma_n)|/n$ is given by*

$$\lim_{n\to\infty}\frac{|\mathcal{P}'_n(W,\gamma_n)|}{n} = \xi(1-\mathcal{C}(W)) \quad (85)$$

In Corollary 19 and Proposition 20, we specialize the general result of Theorem 18 to our needs.

**Corollary 19.** *Let $W$ be an arbitrary BSM channel. Then for any positive constant $\beta < \tfrac{1}{2}$ we have*

$$\lim_{n\to\infty}\frac{|\mathcal{P}'_n(W,2^{-n^\beta})|}{n} = 1 - \mathcal{C}(W) \quad (86)$$

*Proof.* Applying the Stirling formula to both sides of (83), it can be shown (cf. [15]) that

$$a(m,\xi) = \frac{m}{2} + \frac{Q^{-1}(\xi)}{2}\sqrt{m} + o(\sqrt{m})$$

where $Q(x)$ is the probability that a standard normal random variable will obtain a value larger than $x$. Consequently, for all positive $\xi < 1$, there exists a positive constant $c_\xi$ such that

$$a(m,\xi) \geqslant \frac{m}{2} - c_\xi \sqrt{m}$$

for all sufficiently large $m$. This implies that for any positive constant $\beta < \tfrac{1}{2}$, any function $f(m)$ from the positive integers to the interval $(0,1)$, and for all sufficiently large $m$, the following inequality holds

$$\alpha(m,\xi)(1+f(m)) > \frac{a(m,\xi)}{m} \geqslant \frac{1}{2} - \frac{c_\xi}{\sqrt{m}} \geqslant \beta$$

This, in turn, implies that for all sufficiently large $n = 2^m$, we have $\gamma_n < 2^{-n^\beta}$ where $\gamma_n$ is the function defined in (84). Therefore, the set $\mathcal{P}'_n(W,\gamma_n)$ is a subset of the set $\mathcal{P}'_n(W,2^{-n^\beta})$ for all sufficiently large $n$.

Let us assume for a moment that the limit on the left-hand side of (86) exists, call it $L$. If so, we can conclude from Theorem 18, along with the fact that $\mathcal{P}'_n(W,\gamma_n) \subset \mathcal{P}'_n(W,2^{-n^\beta})$ for all sufficiently large $n$, that

$$\lim_{n\to\infty}\frac{|\mathcal{P}'_n(W,2^{-n^\beta})|}{n} \geqslant \lim_{n\to\infty}\frac{|\mathcal{P}'_n(W,\gamma_n)|}{n} = \xi(1-\mathcal{C}(W))$$

Since this holds for all positive $\xi < 1$, the lowest possible value of $L$ is $1 - \mathcal{C}(W)$. Now observe that the set $\mathcal{P}'_n(W,2^{-n^\beta})$ and the set $\mathcal{G}_n(W,\beta)$ of good channels, defined in (13), do not intersect. Therefore, for all $n = 2^m$ we have

$$\frac{|\mathcal{P}'_n(W,2^{-n^\beta})|}{n} \leqslant 1 - \frac{|\mathcal{G}_n(W,\beta)|}{n}$$

Together with Theorem 1, this implies that the limit $L$ indeed exists, and is equal to $1 - \mathcal{C}(W)$. ∎

**Proposition 20.** *Let $W$ be an arbitrary BSM channel, and let $\beta < \tfrac{1}{2}$ be a positive constant. Further, let $\mathcal{P}_n(W,\delta_n)$ be the index set of $\delta_n$-poor bit-channels, as defined in (50), and suppose that there exist positive constants $c_1$ and $c_2$ such that*

$$c_1 2^{-n^\beta} \leqslant \delta_n \leqslant 1 - c_2 \quad (87)$$

*for all sufficiently large $n$. Then the asymptotic behavior of the fraction $|\mathcal{P}_n(W,\delta_n)|/n$ is given by*

$$\lim_{n\to\infty}\frac{|\mathcal{P}_n(W,\delta_n)|}{n} = 1 - \mathcal{C}(W) \quad (88)$$

*Proof.* Let $W_1, W_2, \ldots, W_n$ denote the bit-channels, as before. It was shown by Arıkan in [3, Proposition 1] that

$$\mathcal{C}(W_i) \leqslant \sqrt{1 - Z(W_i)^2} \quad (89)$$

for all $i \in [n]$. Fix a positive constant $\alpha$ such that $\beta < \alpha < \tfrac{1}{2}$. Since $Z(W_i) \geqslant 1 - 2^{-n^\alpha}$ for all $i \in \mathcal{P}'_n(W,2^{-n^\alpha})$ by definition, Arıkan's bound (89) implies that

$$\mathcal{C}(W_i) \leqslant 2^{-(n^\alpha - 1)/2} \quad (90)$$

for $i \in \mathcal{P}'_n(W,2^{-n^\alpha})$. For all sufficiently large $n$, the right-hand side of (90) is less than the left-hand side of (87), and therefore

$$\mathcal{P}'_n(W,2^{-n^\alpha}) \subseteq \mathcal{P}_n(W,\delta_n) \quad (91)$$

Also note that the condition $\delta_n \leqslant 1 - c_2$ on the right-hand side of (87) implies that for all sufficiently large $n$, we have

$$\mathcal{P}_n(W,\delta_n) \cap \mathcal{G}_n(W,\beta) = \varnothing \quad (92)$$

The proposition now follows by combining (91) and (92) with Corollary 19 and Theorem 1, respectively. ∎

We are now ready to prove that for a wide range of security functions, our strong-security coding scheme operates at a rate that approaches the secrecy capacity.

**Theorem 21.** *For any security function $\delta_n$ that satisfies (87), the rate $R_n$ of the corresponding strong-security coding scheme approaches the secrecy capacity, namely*

$$\lim_{n \to \infty} R_n = \mathcal{C}(W^*) - \mathcal{C}(W) \qquad (93)$$

*Proof.* A lower bound on the rate $R_n$ is given in (81). Along with Proposition 20 and Theorem 1, this immediately shows that $\lim_{n\to\infty} R_n \geqslant \mathcal{C}(W^*) - \mathcal{C}(W)$. In order to establish *equality* in (93), let us partition the set $\mathcal{R} = [n] \setminus \mathcal{P}_n(W, \delta_n)$ in (51) into two subsets, as in Figure 2. These subsets are defined as follows:

$$\mathcal{X} \stackrel{\text{def}}{=} \mathcal{R} \cap \mathcal{B}_n(W^*, \beta) \qquad (94)$$

$$\mathcal{Y} \stackrel{\text{def}}{=} \mathcal{R} \cap \mathcal{G}_n(W^*, \beta) \qquad (95)$$

Note that the set $\mathcal{X}$ in (94) and the set $\mathcal{B}$ in (53) form a partition of $\mathcal{B}_n(W^*, \beta)$, as illustrated in Figure 2. It follows that

$$R_n = \frac{|\mathcal{P}_n(W, \delta_n)|}{n} - \frac{|\mathcal{B}_n(W^*, \beta)|}{n} + \frac{|\mathcal{X}|}{n} \qquad (96)$$

We will show in Proposition 22 of the next subsection that the fraction $|\mathcal{X}|/n$ vanishes as $n \to \infty$. With this, the theorem follows from (96), Proposition 20, and Theorem 1. ∎

### F. Reliability of the Strong-Security Coding Scheme

If the main channel $W^*$ is noiseless, Bob can trivially recover the message with probability 1 as follows. Bob receives the vector $\boldsymbol{Y} = \boldsymbol{V} G_n$. Since the Arıkan transform matrix $G_n = P_n G^{\otimes m}$ is its own inverse over $\mathbb{F}_2$, Bob can compute $\boldsymbol{V} = \boldsymbol{Y} G_n$ and set $\widehat{\boldsymbol{U}} = \boldsymbol{V}_{\mathcal{A}}$. Clearly $\widehat{\boldsymbol{U}} = \boldsymbol{U}$, and condition (1) is trivially satisfied.

What happens if the main channel $W^*$ is not noiseless? Suppose that Bob attempts to use the successive cancellation decoder, as in Section V. Then, according to Theorem 2 and Proposition 3, Bob's probability of error is upper-bounded by the sum of the Bhattacharyya parameters $Z(W_i^*)$ of those bit-channels that are not fixed (to zero). The index set of the bit-channels that are not fixed in our strong-security coding scheme is given by

$$\mathcal{A} \cup \mathcal{R} = \mathcal{G}_n(W^*, \beta) \cup \mathcal{X}$$

where $\mathcal{X}$ is the set defined in (94). The sum of the Bhattacharya parameters $Z(W_i^*)$ over the set $\mathcal{G}_n(W^*, \beta)$ is bounded by $2^{-n^\beta}$ by definition, and therefore

$$\Pr\{\widehat{\boldsymbol{U}} \neq \boldsymbol{U}\} \leqslant 2^{-n^\beta} + \sum_{i \in \mathcal{X}} Z(W_i^*) \qquad (97)$$

Unfortunately, we are not aware of any useful bounds on the sum $\sum_{i \in \mathcal{X}} Z(W_i^*)$. Thus we do not have a proof that the strong-security coding scheme satisfies the reliability condition (1).

Observe, however, that the security and reliability requirements are fundamentally different. Whether we're interested in the theory or in the practice of wiretap channels, security always requires a proof. It cannot be established through a computational procedure, such as simulation. On the other hand, re-liability can be (and often is) established through computation in practice. For example, it is very common to rely on simulations to verify the reliability performance of error-correcting codes. We believe that, in practice, our strong-security coding scheme can be decoded to achieve low probabilities of error.

First, the following proposition shows that if $W$ is degraded with respect to $W^*$, then the set $\mathcal{X}$ that gives us trouble in successive cancellation decoding is small.

**Proposition 22.** *Let $\mathcal{X}$ be the index set of bit-channels that are not $\delta_n$-poor for Eve yet bad for Bob, as defined in (94). Then*

$$\lim_{n \to \infty} \frac{|\mathcal{X}|}{n} = 0$$

*for any security function $\delta_n$ that satisfies (87), provided Eve's channel $W$ is degraded with respect to Bob's channel $W^*$.*

*Proof.* We claim that the sets $\mathcal{X}$, $\mathcal{G}_n(W^*, \beta)$, and $\mathcal{P}_n(W^*, \delta_n)$ are pairwise disjoint, and therefore

$$\frac{|\mathcal{X}|}{n} + \frac{|\mathcal{G}_n(W^*, \beta)|}{n} + \frac{|\mathcal{P}_n(W^*, \delta_n)|}{n} \leqslant 1 \qquad (98)$$

Since $\mathcal{X} \subseteq \mathcal{B}_n(W^*, \beta)$ by definition, and $\mathcal{B}_n(W^*, \beta)$ is the complement of $\mathcal{G}_n(W^*, \beta)$, it is clear that $\mathcal{X} \cap \mathcal{G}_n(W^*, \beta) = \varnothing$. It is also clear that $\mathcal{P}_n(W^*, \delta_n) \cap \mathcal{G}_n(W^*, \beta) = \varnothing$, as in (92). Furthermore, since $\mathcal{X} \subseteq \mathcal{R}$ and $\mathcal{R} = [n] \setminus \mathcal{P}_n(W, \delta_n)$ by definition, we have $\mathcal{X} \cap \mathcal{P}_n(W, \delta_n) = \varnothing$. Consequently, in order to prove our claim in (98), it would suffice to show that

$$\mathcal{P}_n(W^*, \delta_n) \subseteq \mathcal{P}_n(W, \delta_n)$$

But this follows immediately from Lemma 4 along with the definition of the set of $\delta_n$-poor channels in (50). Now observe that as $n \to \infty$, the fraction $|\mathcal{G}_n(W^*, \beta)|/n$ converges to the capacity $\mathcal{C}(W^*)$ by Theorem 1, whereas the fraction $|\mathcal{P}_n(W^*, \delta_n)|/n$ converges to $1 - \mathcal{C}(W^*)$ by Proposition 20. Together with (98), this completes the proof of the proposition. ∎

Depending upon how small the set $\mathcal{X}$ turns out to be in practice, various decoding solutions are potentially applicable.

First, it is quite possible that $\mathcal{X} = \varnothing$ in many situations, especially if the main channel is much better than the wiretap channel. In this case, successive cancellation decoding can be used "as is," and Bob's probability of error is at most $2^{-n^\beta}$.

The following example illustrates this situation. In order to obtain the numerical values given in this example, we have used the methods of [37] to evaluate the polar bit-channels.

**Example.** Suppose that both the main channel and the wiretap channel are binary symmetric channels, say $C_1 = \text{BSC}(p_1)$ and $C_2 = \text{BSC}(p_2)$. Let us further assume that $p_1 = 10^{-3}$ and that the error-rate required at the output of the main-channel decoder is $10^{-9}$. This is often the case in optical fiber communications [31]. We will use the polar transformation (5) of length $n = 2^{20}$. Indeed, codes of this length are already in use *today* in proprietary 100 GbE fiber-optic systems. We also adopt the following stringent security criterion: we require that the mutual information $I(\boldsymbol{U}; \boldsymbol{Z})$ between messages at the input to our encoder and observations at the output of the wiretap channel is less than $10^{-30}$. Using the methods developed in this section, we can simultaneously guarantee reliability of $10^{-9}$ and secu-rity of $10^{-30}$ at communication rates close to the secrecy capacity. The following table:

| $p_2$ | Rate $R$ | % of $\mathcal{C}_s$ |
|---|---|---|
| 0.45 | 0.933 | 95.1% |
| 0.40 | 0.882 | 91.9% |
| 0.35 | 0.817 | 88.5% |
| 0.30 | 0.738 | 84.8% |
| 0.25 | 0.647 | 80.9% |
| 0.20 | 0.543 | 76.4% |
| 0.15 | 0.425 | 71.1% |
| 0.10 | 0.293 | 64.0% |

(99)

summarizes these rates as a function of the bit error-rate $p_2$ of the wiretap channel. The third column in the table gives the ratio $R/\mathcal{C}_s$, expressed as a percentage. Notably, in *all of these cases*, we have $\mathcal{X} = \varnothing$. In fact, the set $\mathcal{X}$ remains empty until the bit error-rate on the wiretap channel decreases down to $p_2 = 0.066$, in which case $|\mathcal{X}| = 1$. But the secrecy capacity for $p_2 \leqslant 0.066$ is less than 0.3395, which is probably too small to be of practical interest (in fiber-optic communications). □

Now suppose that the set $\mathcal{X}$ is nonempty. Observe that this set is fixed and known a priori to all the parties (Alice, Bob, and Eve). In successive cancellation decoding, Bob makes his decisions $\widehat{v}_1, \widehat{v}_2, \ldots, \widehat{v}_n$ sequentially using the decision rule (12). Bob will know a priori that this decision rule is unreliable whenever an index $i \in \mathcal{X}$ is reached. Therefore, Bob could follow *both* alternatives $\widehat{v}_i = 0$ and $\widehat{v}_i = 1$ for all $i \in \mathcal{X}$. Doing so increases the decoding complexity by a factor of $2^{|\mathcal{X}|}$. But if $|\mathcal{X}|$ is a small constant (say, $\mathcal{X}$ contains only a couple of bit-channels), this is not unreasonable.

What can Bob do if the set $\mathcal{X}$ is larger? It is well known that in successive cancellation decoding, a single incorrect decision affects all the following decisions, making them unreliable. We propose to *take advantage* of this phenomenon in order to reduce the decoding complexity. Let $i_1$ be the smallest index in $\mathcal{X}$. Suppose that upon branching with $\widehat{v}_{i_1} = 0$ and $\widehat{v}_{i_1} = 1$, the decoder begins to compute the channel noise (e.g. the Hamming distance to the received vector on a BSC channel) accumulated along each of the two decision paths being followed. Due to the "error propagation" induced by the incorrect decision at $i_1$, we expect this estimated noise to accumulate rapidly along the incorrect path. On the other hand, along the correct path, the channel noise should accumulate slowly, governed by the statistics of $W^*$ that are known a priori. This means that the decoder can detect, with high probability, which of the two paths being followed is incorrect. Once the decoder finds that the path with the higher accumulated noise is sufficiently unlikely, according to the channel statistics, this path can be safely discarded.

Of course, it is possible that the second smallest index $i_2 \in \mathcal{X}$ is reached before one of the two paths opened at $i_1 \in \mathcal{X}$ can be discarded. In this case, the decoder would need to begin following four paths. Once the third index $i_3 \in \mathcal{X}$ is reached, the decoder might be following 1, 2, 3, or 4 paths. And so on. In practice, one could design the decoder to follow at most $M$ paths, where $M$ is a pre-determined limit dictated by the decoder complexity considerations. The situation is quite similar to decision-feedback equalization on ISI channels using a bank of $M$ zero-forcing DFEs. That scenario was analyzed in [42], where it is shown that error-propagation caused by incorrect decisions can be used to discard erroneous decision-feedback paths. It is also shown in [42] that, in practice, small values of $M$ often suffice to achieve very good performance.

We have limited our consideration herein to successive cancellation decoding, or variants thereof. It is also possible that other methods of decoding polar codes (such as belief propagation [17] or recursive-list decoding [12]) may be relatively robust to not fixing a small set $\mathcal{X}$ of bad channels. Analysis of such decoders is a research problem of independent interest.

## VII. DISCUSSION AND OPEN PROBLEMS

We briefly mention certain straightforward extensions of our results. So far, we have considered exclusively binary-input symmetric wiretap channels. However, it is well known that the polarization phenomenon extends to other types of channels. Arıkan shows in [3] that, given a non-symmetric binary-input channel $W$, polar codes achieve its symmetric capacity $\mathcal{I}(W)$ in the *average sense*. This implies that our coding scheme achieves the symmetric capacity difference $\mathcal{I}(W^*) - \mathcal{I}(W)$, also in the average sense. Specifically, suppose we modify our encoding algorithm in Section IV as follows. Instead of constructing the vector $\boldsymbol{v} \in \{0,1\}^n$ by setting $\boldsymbol{v}_\mathcal{R} = \boldsymbol{e}$, $\boldsymbol{v}_\mathcal{A} = \boldsymbol{u}$, and $\boldsymbol{v}_\mathcal{B} = \boldsymbol{0}$, we set $\boldsymbol{v}_\mathcal{R} = \boldsymbol{e}$, $\boldsymbol{v}_\mathcal{A} = \boldsymbol{u}$, and $\boldsymbol{v}_\mathcal{B} = \boldsymbol{s}$, where $\boldsymbol{s}$ is a fixed binary vector known a priori to all the parties. Then there exists *some choice* of $\boldsymbol{s}$ such that Theorem 8 and Theorem 9 hold. Based upon the results of Arıkan in [3], exactly the same proof as before (Lemma 6 and Lemma 7) applies.

Our results also extend to discrete memoryless channels with non-binary input. It was recently proved in [34] that channels with an input alphabet of *prime size* $q$ are polarized by the same transformation (5), and the corresponding versions of Theorem 1 and Theorem 2 hold. The probability of error under successive cancellation decoding still scales as $O(2^{-n^\beta})$ for all prime $q$. This means that our proof of Lemmas 6 and 7 goes through essentially "as is" (as long as $r$ is replaced by $r \log_2 q$ throughout). If the size $q$ of the input alphabet is not prime, polarization requires either a randomized permutation on the input or multilevel coding (see [34] for more details). It can be shown that our results in Section V extend to this case as well.

It is not clear whether the strong-security results of the previous section can be similarly extended to non-symmetric and/or to non-binary-input wiretap channels. We believe they can, and pose a proof of this as an open problem.

Another open problem of great interest is how to code for the situation where the wiretap channel $W$ in *not* degraded with respect to the main channel $W^*$. Note that channel degradation is sufficient but, to the best of our knowledge, not necessary for our coding scheme to work. What seems to be necessary is that the set of bit-channels that are "good" for Eve but "bad" for Bob is either empty (as in Section V) or at least very small (as in Section VI). Unfortunately, in the general case, there is no reason why the number of such bit-channels could not be large.

Finally, we point out that all the constructions in this paper are only as explicit as the polar codes themselves. An exact algorithm for computing the sets $\mathcal{G}_n(W, \beta)$ and $\mathcal{P}_n(W, \delta)$ in (13) and (50) was given by Arıkan in [3]. However, this algorithm requires time and memory that grow exponentially with the code length $n$. Since then, several heuristic algorithms for this problem have been proposed [2, 29, 30]. However, these algorithms do not provide useful guarantees on the quality of their output. Such guarantees are clearly essential to establish the security of our coding scheme. Fortunately, the problem has been resolved in [37]. The algorithm of [37] runs in linear time, and makes it possible to compute upper and lower bounds on the capacity of polar bit-channels with an arbitrary degree of precision. For example, to establish the results reported in (99), we have run the algorithm of [37] with a precision of 300 bits (which is necessary to provide meaningful guarantees of security down to $10^{-30}$).

### Acknowledgment

We thank Mihir Bellare, Hamed Hassani, Satish Korada, Ryutaroh Matsumoto, Ido Tal, Emre Telatar, and Rüdiger Urbanke for very helpful discussions. We are especially grateful to Satish Korada for pointing us to Lemma 4.7 of his Ph.D. thesis, and to Ido Tal for his help with the example in Section VI-F.

### References

[1] M. Andersson, V. Rathi, R. Thobaben, J. Kliewer, and M. Skoglund, "Nested polar codes for wiretap and relay channels," preprint of June 17, 2010, arxiv.org/abs/1006.3573.

[2] E. Arıkan, "A performance comparison of polar codes and Reed-Muller codes," *IEEE Comm. Letters*, vol. 12, no. 6, pp. 447-449, June 2008.

[3] E. Arıkan, "Channel polarization: A method for constructing capacity-achieving codes for symmetric binary-input memoryless channels," *IEEE Trans. Inform. Theory*, vol. 55, no. 7, pp. 3051-3073, July 2009.

[4] E. Arıkan and E. Telatar, "On the rate of channel polarization," preprint of July 24, 2008, arxiv.org/abs/0807.3806.

[5] M. Bellare, A. Desai, E. Jokipii, and Ph. Rogaway, "A concrete security treatment of symmetric encryption," *Proc. IEEE Symp. Foundations Computer Science*, pp. 394–403, Miami Beach, FL., October 1997.

[6] C. Bennett, G. Brassard, C. Crépeau, and U.M. Maurer, "Generalized privacy amplification," *IEEE Trans. Inform. Theory,* vol. 41, no. 6, pp. 1915-1923, November 1995.

[7] M. Cheraghchi, F. Didier, and A. Shokrollahi, "Invertible extractors and wiretap protocols," preprint of January 14, 2009, arxiv.org/abs/0901.2120.

[8] G. Cohen and G. Zémor, "The wire-tap channel applied to biometrics," *Proc. Intern. Symp. Information Theory and its Applications*, Parma, Italy, October 2004.

[9] G. Cohen and G. Zémor, "Syndrome-coding for the wiretap channel revisited," *Proc. IEEE Information Theory Workshop*, pp. 33-36, Chengdu, China, October 2006.

[10] T.M. Cover and J.A. Thomas, *Elements of Information Theory*. 2nd Ed., Hoboken, NJ: John Wiley & Sons, 2006.

[11] I. Csiszár and J. Körner, "Broadcast channels with confidential messages," *IEEE Trans. Inform. Theory,* vol. 24, no. 3, pp. 339–348, May 1978.

[12] I. Dumer and K. Shabunov, "Soft-decision decoding of Reed-Muller codes: recursive lists," *IEEE Trans. Inform. Theory,* vol. 52, no. 3, pp. 1260–1266, March 2006.

[13] R.G. Gallager, *Information Theory and Reliable Communication*. Hoboken, NJ: John Wiley & Sons, 1968.

[14] S. Goldwasser and S. Micali, "Probabilistic encryption," *J. Comput. Syst. Sci.* vol. 28, pp. 270–299, April 1984.

[15] S. Hassani and R.L. Urbanke, "On the scaling of polar codes: The behavior of polarized channels," *Proc. IEEE Intern. Symp. Information Theory*, pp. 874–878, Austin, TX., June 2010.

[16] E. Hof and S. Shamai, "Secrecy-achieving polar-coding for binary-input memoryless symmetric wire-tap channels," *IEEE Trans. Inform. Theory*, submitted for publication, May 2010.

[17] N. Hussami, R.L. Urbanke, and S.B. Korada, "Performance of polar codes for channel and source coding," *Proc. IEEE Intern. Symp. Information Theory*, pp. 1488–1492, Seoul, Korea, June 2009.

[18] S.B. Korada, *Polar Codes for Channel and Source Coding*, Ph.D. dissertation, EPFL, Lausanne, Switzerland, May 2009.

[19] S. B. Korada, E. Şaşoğlu, and R.L. Urbanke, "Polar codes: Characterization of exponent, bounds, and constructions," *Proc. IEEE Intern. Symp. Information Theory*, pp. 1483–1487, Seoul, Korea, June 2009.

[20] O.O. Koyluoğlu and H. El Gamal, "Polar coding for secure transmission and key agreement," *Proc. IEEE Intern. Symp. Personal Indoor and Mobile Radio Comm.*, pp. 2698–2703, Istanbul, Turkey, September 2010.

[21] S. Leung-Yan-Cheong, "On a special class of wire-tap channels," *IEEE Trans. Inform. Theory,* vol. 23, no. 5, pp. 625–627, September 1977.

[22] S. Leung-Yan-Cheong and M.E. Hellman, "The Gaussian wire-tap channel," *IEEE Trans. Inform. Theory,* vol. 24, no. 4, pp. 451–456, July 1978.

[23] Y. Liang, H.V. Poor, and S. Shamai, "Information theoretic security," *Foundations and Trends in Communications and Information Theory*, vol. 5, issue 4-5, pp. 355–580, 2008.

[24] H. Mahdavifar and A. Vardy, "Achieving the secrecy capacity of wiretap channels using polar codes," preprint of December 31, 2009, arxiv.org/abs/1001.0210.

[25] H. Mahdavifar and A. Vardy, "Achieving the secrecy capacity of wiretap channels using polar codes," *Proc. IEEE Intern. Symp. Information Theory*, pp. 913–917, Austin, TX., June 2010.

[26] R. Matsumoto, "On the randomness in the encoder," private communication, June 2010.

[27] U.M. Maurer, "The strong secret key rate of discrete random triples," pp. 271-285 in *Communication and Cryptography — Two Sides of One Tapestry*, R.E. Blahut et al. (Eds.), Boston: Kluwer Academic, 1994.

[28] U.M. Maurer and S. Wolf, "Information-theoretic key agreement: From weak to strong secrecy for free," *Lect. Notes Computer Science,* vol. 1807, pp. 351–368, Springer-Verlag, 2000.

[29] R. Mori and T. Tanaka, "Performance and construction of polar codes on symmetric binary-input memoryless channels," *Proc. IEEE Intern. Symp. Information Theory*, pp. 1496–1500, Seoul, Korea, June 2009.

[30] R. Mori and T. Tanaka, "Performance of polar codes with the construction using density evolution," *IEEE Comm. Letters*, vol. 13, no. 7, pp. 519–521, July 2009.

[31] D.K. Mynbaev and L.L. Scheiner, *Fiber-Optic Communications Technology*. Englewood Cliffs, NJ: Prentice Hall, 2000.

[32] L.H. Ozarow and A.D. Wyner, "Wire-tap channel II," *Bell Lab. Tech. J.,* vol. 63, no. 10, pp. 2135–2157, December 1984.

[33] Y. Polyanskiy, H.V. Poor, and S. Verdú, "Channel coding rate in the finite blocklength regime," *IEEE Trans. Inform. Theory,* vol. 56, no. 5, pp. 2307-2359, May 2010.

[34] E. Şaşoğlu, E. Telatar, and E. Arıkan "Polarization for arbitrary discrete memoryless channels," *Proc. IEEE Information Theory Workshop*, pp. 144–148, Taormina, Italy, October 2009.

[35] V. Strassen, "Asymptotische Abschätzungen in Shannon's Informationstheorie," *Trans. 3-rd Prague Conf. Information Theory, Statistical Decision Functions, and Random Processes*, pp. 689–732, Publ. Czech. Acad. Sci., Prague, 1964.

[36] A. Suresh, A. Subramanian, A. Thangaraj, M. Bloch, and S.W. McLaughlin, "Strong secrecy for erasure wiretap channels," *Proc. IEEE Information Theory Workshop*, Dublin, Ireland, September 2010.

[37] I. Tal and A. Vardy, "How to construct polar codes," presented at the *IEEE Information Theory Workshop*, Dublin, Ireland, September 2010.

[38] A. Thangaraj, S. Dihidar, A.R. Calderbank, S.W. McLaughlin, and J. Merolla, "Applications of LDPC codes to the wiretap channel," *IEEE Trans. Inform. Theory,* vol. 53, no. 8, pp. 2933-2945, August 2007.

[39] M. Van Dijk, "On a special class of broadcast channels with confidential messages," *IEEE Trans. Inform. Theory,* vol. 43, no. 2, pp. 712–714, March 1997.

[40] V.K. Wei, "Generalized Hamming weights for linear codes," *IEEE Trans. Inform. Theory,* vol. 37, no. 5, pp. 1412-1418, September 1991.

[41] A.D. Wyner, "The wire-tap channel," *Bell System Tech. J.,* vol. 54, no. 8, pp. 1355–1387, October 1975.

[42] D. Yellin, A. Vardy, and O. Amrani, "Joint equalization and coding for intersymbol interference channels," *IEEE Trans. Inform. Theory*, vol. 43, no. 2, pp. 409–425, March 1997.